\documentclass[12pt,draftclsnofoot,onecolumn]{IEEEtran}
\usepackage{latexsym}
\usepackage{amsmath}
\usepackage{url}
\usepackage{amsfonts}
\usepackage{graphicx}
\usepackage{color}
\usepackage{bbold}
\usepackage{caption}
\usepackage{subcaption}
\usepackage{url}
\usepackage{upgreek}
\usepackage{epstopdf}
\usepackage{cite}
\usepackage{cleveref}

\usepackage{bm}        
\usepackage{algorithm}
\usepackage{algpseudocode}

\errorcontextlines\maxdimen

\usepackage{etoolbox}
\usepackage{tikz}
\usetikzlibrary{tikzmark}
\usetikzlibrary{calc}

\newcommand{\ALGtikzmarkcolor}{black}
\newcommand{\ALGtikzmarkextraindent}{4pt}
\newcommand{\ALGtikzmarkverticaloffsetstart}{-.5ex}
\newcommand{\ALGtikzmarkverticaloffsetend}{-.5ex}
\makeatletter
\newcounter{ALG@tikzmark@tempcnta}

\newcommand\ALG@tikzmark@start{%
    \global\let\ALG@tikzmark@last\ALG@tikzmark@starttext%
    \expandafter\edef\csname ALG@tikzmark@\theALG@nested\endcsname{\theALG@tikzmark@tempcnta}%
    \tikzmark{ALG@tikzmark@start@\csname ALG@tikzmark@\theALG@nested\endcsname}%
    \addtocounter{ALG@tikzmark@tempcnta}{1}%
}

\def\ALG@tikzmark@starttext{start}
\newcommand\ALG@tikzmark@end{%
    \ifx\ALG@tikzmark@last\ALG@tikzmark@starttext
    \else
        \tikzmark{ALG@tikzmark@end@\csname ALG@tikzmark@\theALG@nested\endcsname}%
        \tikz[overlay,remember picture] \draw[\ALGtikzmarkcolor] let \p{S}=($(pic cs:ALG@tikzmark@start@\csname ALG@tikzmark@\theALG@nested\endcsname)+(\ALGtikzmarkextraindent,\ALGtikzmarkverticaloffsetstart)$), \p{E}=($(pic cs:ALG@tikzmark@end@\csname ALG@tikzmark@\theALG@nested\endcsname)+(\ALGtikzmarkextraindent,\ALGtikzmarkverticaloffsetend)$) in (\x{S},\y{S})--(\x{S},\y{E});%
    \fi
    \gdef\ALG@tikzmark@last{end}%
}

\apptocmd{\ALG@beginblock}{\ALG@tikzmark@start}{}{\errmessage{failed to patch}}
\pretocmd{\ALG@endblock}{\ALG@tikzmark@end}{}{\errmessage{failed to patch}}
\makeatother

\DeclareMathOperator*{\argmin}{arg\,min}  
\DeclareMathOperator*{\argmax}{arg\,max}

\newcommand{\e}{\mathrm{e}} 
\newcommand{\eqdef}{\stackrel{\textrm{\tiny def}}{=}}

\newcommand{\y}{\bm{y}}
\newcommand*{\punto}{\makebox[1ex]{\textbf{$\cdot$}}}%

\newcommand{\ed}{\color{black}} 

\usepackage[font=small]{caption}

\begin{document}

\title{\ed Mobile Positioning in Multipath Environments: a Pseudo Maximum Likelihood approach}

\author{Alessio Fascista, Angelo Coluccia, \IEEEmembership{Senior Member, IEEE}, and Giuseppe Ricci, \IEEEmembership{Senior Member, IEEE} 
\thanks{All authors are with the Dipartimento di Ingegneria dell'Innovazione, Universit\`a del Salento, via Monteroni, 73100 Lecce, Italy. E-Mail: name.surname@unisalento.it.}
}

\maketitle

\begin{abstract}
  The problem of {\ed mobile position estimation in multipath} scenarios is addressed. A low-complexity, fully-adaptive algorithm is proposed, based on the {\ed pseudo maximum likelihood approach}. The processing is done exclusively on-board at the mobile node by exploiting {\ed narrowband downlink radio signals}.
The proposed algorithm is able to estimate via adaptive beamforming (with spatial smoothing) the optimal projection {\ed matrices that maximize the likelihood; in addition,  it can associate the line-of-sight over the trajectory}, hence achieving an integration gain.   
The performance assessment shows that the proposed algorithm is very effective in (even severe) multipath conditions, outperforming natural competitors also when the number of antennas and snapshots is kept at the theoretical minimum. 
\end{abstract}


\section{Introduction}\label{sec:introduction}
 

  \IEEEPARstart{P}{osition}  estimation is important in many contexts such as wireless sensor networks, vehicular scenarios, and for navigation/tracking at large. 
Locating a node in a wireless system  involves  radio signals propagating between the node and a number of base stations (BSs) at known positions.
Different information can be exploited, namely  received signal strength (RSS), time (difference) of arrival (TOA/TDOA), and angle of arrival (AOA)  \cite{Patwari_SPM,Tomic1,CCFR,Tomic2}. 
Techniques based on the RSS, although   simpler, are not able to provide sufficient location accuracy due to the great variability of the power in wireless channels, especially in case of strong multipath.  On the other hand, time of arrival (TOA) \cite{LS_ToA}
or time difference of arrival (TDOA) \cite{Savvides,TASSP_TDoA} techniques are challenging in terms of
clock synchronization and are very sensitive to multipath.

  {\ed AOA-based methods, traditionally more linked to surveillance (e.g., radar) and related fields, are currently experiencing renewed interest due to the widespread of MIMO technologies in 4G cellular networks}, and are becoming even more attractive for  5G mmWave scenarios in which array size significantly shrinks, thus allowing integration in mobile terminals (smartphones) \cite{MIMOsmartphone, Shahmansoori_5GLocalization}. This technological evolution in cellular communications, together with the wide availability of sensing modules (e.g., kinematic sensors such as INS) and computing capabilities in modern mobile devices, is paving the way for innovative localization paradigms.
 {\ed Moreover, location awareness is very important for autonomous vehicles, robotics, and other vehicular applications, in which the same technological innovations are becoming progressively available.
 
Motivated by the above considerations, we address a localization setup in which downlink  signals   from  one or more  BSs are exploited by a  mobile  node  equipped with an antenna array, to estimate its own position. Differently from the more conventional  uplink  setup ---  in which  BSs  receive  the  signal from the node at unknown position, perform some processing (e.g., AOA estimation), and then send such a local information to a central node for position estimation --- in the considered downlink scenario the whole procedure is performed at the mobile node by leveraging  broadcast radio signals, also exploiting the availability of velocity estimates from an onboard sensor. This has the advantage of not requiring further communications, considerably decreasing the bandwidth consumption \cite{fu}. Moreover, antenna arrays are not needed on BSs since the latter do not play any active role in the estimation task.  
 
For this localization setup, we address more specifically the problem of localizing a mobile node while taking explicitly into account the multipath structure. 
As better discussed in Sec. II,
 this is a more challenging goal compared to the typical scenario addressed in the AOA-based localization literature, where multipath is only regarded as a (stochastic) disturbance and a single snapshot of the environment is considered, i.e., a static scenario as in a ``still frame''.
Typically, the localization of a mobile object is instead recast as the problem of tracking its trajectory over time:  in particular, in the celebrated   Kalman  filter  (KF), a  Bayesian  (MMSE)  estimation  approach is adopted in which the  current  position  estimate  is  updated  recursively  in weighted combination with a new (noisy) position estimate (or other position-related information in the extended KF, such as range \cite{Valaee}  or  AOA \cite{CCFR_ITS}, and possibly also velocity measurements),  taking  into  account  the  constraints  induced  by  a  chosen  kinematic  model  (e.g.,  nearly-constant velocity model \cite{Bar-Shalom}).
In this paper, conversely, we process a batch  of  array signals  received  in  previous time  instants, which thus experience the effects of mobility, and devise a \emph{pseudo maximum likelihood}  position estimator. In such an approach, quoting from \cite{Gong},
nuisance parameters are eliminated by ``replacing them by estimates and solving a reduced system of likelihood equations. The method is a reasonable one in problems in which lower dimensional maximum likelihood estimation is feasible while higher dimensional maximum likelihood estimation is intractable''.\footnote{\ed For the sake of completeness, we highlight the difference between pseudo ML and quasi-ML: in the former the likelihood function and the estimation steps are the same as the true ML,  the only difference being that some nuisance parameters are not estimated in the ML sense but with a different technique; conversely, in quasi-ML approaches it is the likelihood function that is approximated or relaxed in some way  before performing the maximization with respect to all unknown parameters.}
For the problem considered in this paper, the  many unknown nuisance parameters at play in multipath environments cannot be estimated in the maximum likelihood (ML) sense; following the pseudo ML rationale, we use different estimates instead: in particular, as better explained later, a combination of spatial smoothing and adaptive beamforming allows us to obtain estimates of the projection matrices (where nuisance parameters ultimately appear) that maximize the likelihood.  
Moreover, mobility is exploited in the localization task to reduce the computational complexity involved in the final position estimation and, at the same time, introduce an integration gain that is beneficial to the ultimate localization accuracy.
 It is worth highlighting that the  proposed algorithm
 does not assume any particular mobility model, i.e., it can be applied in general irrespective of the actual trajectory of the mobile terminal.
 }
 
 The rest of the paper is organized as follows. {\ed In Sec. II we discuss the related work, analyzing in details the novelty of our approach compared to the literature.} In Sec. \ref{sec:sysmodel} we introduce the system model and describe the reference scenario.  In Sec. \ref{sec::DPEderiv} we formulate the estimation problem and illustrate in details the design and derivation of the proposed  pseudo ML algorithm; {\ed we also derive two natural competitors that can be seen as extensions of state-of-the-art approaches. Then, in Sec. \ref{sec:simulation}, we assess} the performance by means of Monte Carlo simulations in  realistic scenarios.  We conclude  in Sec. \ref{sec:conclusions}.
 
 \section{Related work}\label{sec:relatedwork}

Localization approaches can be either direct or indirect.
{\ed In indirect position estimation (IPE) techniques a suboptimal two-step procedure is followed: in the first step, some position-related information is obtained, namely distance or angle estimates; in the second step, such estimates are combined together to obtain the unknown position} \cite{Gholami}. Although popular due to their reduced complexity (e.g., for low-cost WSNs \cite{ILS}),  their accuracy is usually limited, and also some bias may be introduced by the first estimation step  \cite{bias_TWC}. Direct position estimation (DPE) techniques, conversely,  use a single-step  approach to estimate the location directly from the raw signals. In doing so, {\ed the direct link between the collected measurements and the node position is exploited}, resulting in a significant improvement of the achievable performance especially under multipath propagation \cite{closas1,closas2}. 

{\ed In the literature, the most relevant DPE approaches that adopt antenna arrays consider a static scenario, i.e., a snapshot of the environment in which both the unknown position  and all channel effects are assumed static. Moreover, the multipath is typically considered as a disturbance.
For instance, in \cite{weiss2,weiss1} a single-path scenario is addressed in which the multipath is modeled as additive noise. A least squares estimator is developed, which coincides with the ML estimator under the white Gaussian noise assumption. Such an approach has been then extended to  multiple nodes localization \cite{weiss3}. The minimum-variance distortionless response (MVDR) beamformer is adopted in \cite{weiss4,weiss5} to mitigate the effects of multipath, regarded as a disturbance without an explicit model for it, while again the signal is modeled as single-path and the noise is assumed white Gaussian.} DPE methods tailored to special signals such as orthogonal frequency division multiplexing  (OFDM), cyclostationary signals, and intermittent emissions have also been proposed  \cite{barshalom,weiss6,oispuu}.
%

{\ed A context in which multipath can be exploited to gain additional position-related information is the emerging field of MIMO communications in 5G mmWave cellular networks.}
In particular, massive arrays offer the possibility of precisely estimating the parameters of each individual multipath component thanks to their high angular resolution. 
A ML estimator has been developed in \cite{papakon} for localizing a single node assuming a fixed and known number of multipaths, but without providing an efficient way to compute the estimator. Recently, a novel algorithm called direct source localization (DiSouL) has been proposed  \cite{garcia2}, based on a compressed sensing framework that exploits some channel properties to improve the performance. 
{\ed Such DPE algorithms  assume antenna arrays at both transmit and receive sides, with  BSs  receiving signals from a terminal; 
such data are then sent to a central node for joint processing, thus consuming a significant amount of bandwidth \cite{fu}.
The whole procedure requires that BSs play an active role in the whole process; moreover, again,}  the localization problem is solved in a static case.

{\ed
We highlight in the following the several aspects that make our paper significantly different from the related work reviewed above.
\begin{itemize}
    \item[{i)}] The considered localization problem is not the same, because is motivated by a different scenario. In particular, the approaches discussed above focus on a classical uplink setup, i.e., BSs equipped with antenna arrays receiving the signal transmitted from the mobile node; then such data need to be sent to a central node for joint processing. In our case, it is the other way round: the mobile terminal receives (through an array) the downlink signals transmitted from the BSs (which do not need arrays) and the processing is done at the mobile node without any further communication required, also exploiting the availability of velocity estimates from an onboard sensor. Such a scenario is less investigated yet of great interest nowadays, as explained in Sec. I.
    \item[{ii)}] In the related work, the standard approach to localization  is to consider a static scenario (single snapshot of the environment); conversely, we exploit mobility to obtain an   integration gain from the signals collected in past time instants.
    This requires a paradigm shift compared to the static case. First, it is not possible to sample the signal arbitrarily, since after a certain time both position and channel parameters cannot be  assumed stationary anymore (because of the motion): indeed, we sample the output of the matched filter at the exact rate that ensures the noise remains white, and we have carefully established (the details are in the Appendix) the maximum value of the number of samples that is compatible with the coherence assumption. 
    Second, localization of a mobile node implies a domain that grows with time, considering all possible directions where the the motion can take place.
    In the proposed approach, conversely, the search space does not increase with time; just the opposite, it shrinks thanks to a carefully designed strategy in which unlikely points are progressively discarded based on the line-of-sight (LOS) associations over time. 
    \item[{iii)}] The closest approach we could identify in the literature is the DPE formalization in \cite{weiss2, weiss1, weiss4, weiss5}. However, in such papers  the structure of the multipath is completely neglected at the design stage; we conversely adopt a deterministic multipath model, where all channel parameters (directions and complex amplitudes) are  unknowns. This has major implications in the resulting signal processing: in the case of \cite{weiss2, weiss1, weiss4, weiss5} a standard ``matched filter in the angular domain'' (but parameterized in the position) is used, while we need to cope with the many nuisance parameters that cannot be obtained in the ML sense. To estimate them, we first decorrelate the multipath through spatial smoothing, then use an adaptive beamforming approach to estimate directions and amplitudes of the different paths (as better explained in Sec. \ref{sec::DPEderiv}).
\end{itemize}
}

{\ed In summary, the contribution of this paper is a novel pseudo ML approach to localization in multipath scenarios under mobility.} As we will show, the proposed approach has low complexity and is fully adaptive, i.e., it does not require any  tuning or additional information about the environment. To the best of our knowledge, this is the first {\ed general AOA-based localization algorithm that uses only narrowband downlink signals and can cope with multipath while at the same time exploiting mobility}. The processing is done exclusively on-board at the mobile node, {\ed without requiring specific actions at the BSs (which can even have single antennas,} 
as opposed to MIMO scenarios). 
The proposed approach is effective even with a single BS; furthermore, it  outperforms natural competitors also when using a minimal number of antennas and snapshots.
{\ed It can be also used in combination with a tracking algorithm that exploits the obtained position estimates.}

\section{System model}\label{sec:sysmodel}
We consider  $N_{BS}$ BSs located at fixed, known positions and a mobile station (MS) with unknown position.  The MS moves along an arbitrary trajectory, with (generally non-constant) velocity that is measured through an inertial sensor or odometer (as typically available in a vehicle or smartphone). Thus, we will assume at the design stage that velocities are known, but  (noisy) measurements will be used in the implementation. 

The position of the $b$-th BS and of the MS at  time instant $t$ are denoted by $\bm{p}^{b}_{BS} = \left[ x^{b}_{BS} \ y^{b}_{BS}\right]^T$ and $\bm{p}(t) =\left[ x(t) \ y(t)\right]^T$ (where $^T$ is the transpose operator), respectively, where $b \in \mathcal{B}$ and $\mathcal{B} = \{1,2,\ldots,N_{BS}\}$ is the set of univocal BS identifiers. As mentioned, differently from  other localization setups, the BSs are transmit-only (with a single, typically omnidirectional antenna) while the MS is receive-only and equipped with an $M$-element uniform linear antenna array (ULA).\footnote{\ed Since elevation angles  cannot be estimated through an ULA,  2D positions (only azimuth angles) are considered in this paper. 
This is tantamount to considering waves that propagate horizontally; such a condition is realistic in macro-cells where the distance between  transmitter and receiver is  large compared to the height of the antennas, while in other contexts discarding the elevation may introduce an error in the azimuth estimation \cite{elevation}.
The proposed methodology can be extended in principle to address the 3D localization setup; we will discuss this possibility after the derivation of the proposed algorithm in Sec. \ref{sec::DPEderiv}, so that the necessary modifications can be described.} Particularly, each BS  broadcasts a signal with baseband representation $s(t) = \sum_h c_h g(t-hT)$ where $g(\punto)$ denotes a root-raised-cosine (RRCR) signaling pulse known to the receiver, $c_h$s  the transmitted symbols, and $B = (1+\alpha_{\text{\tiny RRCR}})/2T$  the one-sided bandwidth with roll-off factor $\alpha_{\text{\tiny RRCR}} \in [0,1]$.  Notice that the derivations would apply also to the reverse situation in which the MS transmits and the BSs receive; this  is however less attractive since it  requires additional mechanisms to coordinate data collection, including BS synchronization, as discussed.  

The MS executing the  localization algorithm collects and processes the impinging signals coming from the transmitting BSs nearby (assumed in the far field). More specifically, let $\bm{p}_0 = [x_0 \ y_0]^T \eqdef \bm{p}(t_0)$ be the (unknown) MS position at time instant $t_0$ when the localization procedure starts. Moreover, let $t_i, i >0$, denote the time instant at which a signal transmitted by one of the BS has been received; we denote by $b_i \in \mathcal{B}$ the identifier of such a BS.  Since there is a  correspondence between the $i$-th received signal and the transmitting BS $b_i$, in the following we will use only $t_i$ while omitting $b_i$ from the notation.
As concerns the multipath channel, we assume that (i) $T$  is much greater than the channel delay spread $\tau_S$, so that the channel exhibits a constant complex gain response;  (ii) $B$ is much greater than the channel Doppler spread $B_D$ caused by MS mobility, so that the channel response can be assumed to be time-invariant over a small-scale observation period $T_{obs}$. The resulting received signal over a generic time interval $[t_i, t_i + T_{obs}]$ after down-conversion, clock and frequency/phase offsets recovery can be expressed as \cite{rappaport,goldsmith}
\begin{equation}\label{eq::modelflatslow}
\bm{x}(t) = \gamma_i (\bm{x}^{\text{\tiny LOS}}_i + \bm{x}^{\text{\tiny NLOS}}_i) s(t) + \bm{n}(t) \quad t_i \leq t \leq t_i + T_{obs}
\end{equation}
where
\begin{align}
\bm{x}^{\text{\tiny LOS}}_i &= \bm{a}\left(\theta^{\text{\tiny LOS}}_i\right) \label{eq::los}\\ 
\bm{x}^{\text{\tiny NLOS}}_i &= \sum_{m=1}^{D_i} \beta^{m}_i\bm{a}(\theta^{m}_i)  \label{eq::nlos}
\end{align}
in which $\bm{n}(t)$ is  thermal noise and $\bm{a}(\theta)$ is the steering vector representing the array response for a signal impinging with angle $\theta$. 
As to $\gamma_i$ and $\theta^{\text{\tiny LOS}}_i$, they are the complex amplitude coefficient related to large-scale fading (or path-loss) and the AOA of the LOS path at time instant $t_i$, respectively, while $\beta^{m}_i$ and $\theta^{m}_i$ are the complex small-scale fading coefficient and the AOA of the $m$-th multipath component out of  the $D_i$ non-line-of-sight (NLOS) paths. The value of $D_i$ can change at each time instant $t_i$ and is typically unknown; {\ed in the following, we assume a fixed design parameter $D_{\text{max}}$ that can be set (even in a conservative way) based on preliminary considerations\footnote{\ed 
 Classical information-theoretic techniques for model selection can be used, e.g. Akaike's \cite{akaike}, to estimate such a value. In general, some prior knowledge is typically available for a given environment,  based on experiments.}, hence generally differs from the actual $D_i$. 
 As we will show in Sec. \ref{sec:simulation}, where the $D_i$ in the generated signals are mismatched to the fixed design value $D_{\text{max}}$, the proposed approach is robust to misknowledge of such parameters.}

It is worth highlighting that the model above
describes a multipath channel with  \emph{flat} and \emph{slow} fading effects. 
Notice that (i) $T \gg \tau_S$ is tantamount to neglecting the delays $\tau^m_i$s associated to the $D_i$ NLOS paths, that is, $s(t - \tau^m_i) \approx s(t)$ $\forall m$, while (ii) $B \gg B_D$ guarantees that the complex coefficients $\beta^{m}_i$s do not change over the observation period $T_{obs}$. It is also assumed that  $T_{obs}$ is short enough so that the position and velocity of the MS remain approximately constants, i.e., the multipath geometry in terms of $\theta^{\text{\tiny LOS}}_i$  and $\theta^{m}_i$s is unchanged.
The way the parameters $B$ and $T_{obs}$ are chosen will be discussed in the numerical analysis conducted in Sec. \ref{sec:simulation} (and Appendix), where a realistic scenario of MS localization in multipath environments is considered.

For $\bm{n}(t)$ we consider the classical white complex normal model. 
Moreover, assuming a ULA with isotropic antennas (and no mutual coupling), the steering vector  is
\begin{equation}
\bm{a}(\theta) = \left[1 \ e^{j\omega d\sin\theta} \cdots \ e^{j(M-1)\omega d\sin\theta}\right]^T \label{eq:steering}
\end{equation}
where $\omega = 2\pi/\lambda$ represents the incident wave number, $\lambda = c/f_c$ is the carrier wavelength, $f_c$ is the carrier frequency, $c$ is the speed of light, and $d = \lambda/2$ is the ULA interelement spacing.
{\ed Notice that the expression in \eqref{eq:steering} refers to angles $\theta\in (-\pi/2, \pi/2)$ with respect to the normal direction to the array --- i.e., for sensors lying on the $x$-axis, $\theta$ is positive from the $y$-axis (clockwise) in the first quadrant, negative (counterclockwise) in the second quadrant. The radiation pattern of a ULA has a symmetry of revolution around the line where the antennas are located; as a consequence, angles outside $(-\pi/2, \pi/2)$ are anyway ``seen'' as belonging to
such an interval, in particular as the corresponding symmetric angle with respect to the array line. While this introduces an inherent ambiguity for AOA estimation, we will show that the proposed approach overcomes the half-plane limitation of ULAs thanks to a suitably-designed association mechanism.}
Throughout the rest of the paper, we assume that all the parameters $\gamma_i$, $\theta^{\text{\tiny LOS}}_i$, $\beta^m_i$s, $\theta^m_i$s are unknown. 

\begin{figure}
	\centering
	\includegraphics[width=0.45\textwidth]{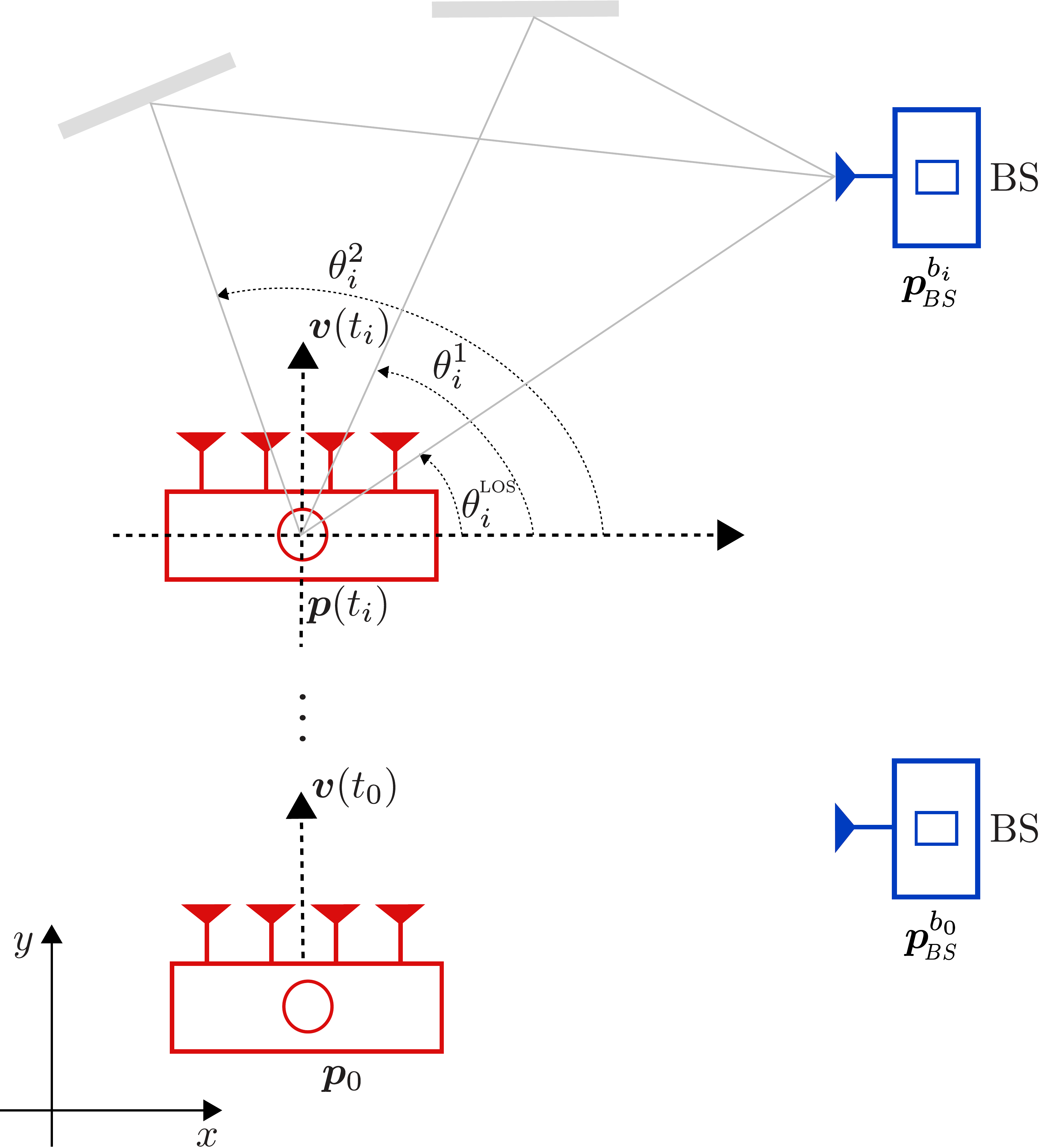}
	\caption{Reference scenario of the considered mobile position estimation problem.}
	\label{fig:ph1}
\end{figure}

At the receiver,  $\bm{x}(t)$ is passed through a matched filter
\begin{equation}
\bm{y}(t) = \int_{t_i}^{t_i + T_{obs}} g^*(\tau - t)\bm{x}(\tau) \mathrm{d}\tau
\end{equation}
(where $^*$ is the complex conjugate operator), whose output is then sampled at a rate $f_s = 1/T$, resulting in the following sequence of received samples\footnote{Without loss of generality,  we consider a signaling pulse with normalized energy, i.e., $\int_{t_i}^{t_i + T_{obs}} |g(\tau)|^2d\tau = 1$.}
\begin{equation}\label{eq::samples}
\bm{y}_{i,n} = \gamma_i (\bm{x}^{\text{\tiny LOS}}_i +  \bm{x}^{\text{\tiny NLOS}}_i)c_{i,n} + \bm{\nu}_{i,n} \quad n=0,\ldots,N-1 
\end{equation}
where $N = \left\lfloor\frac{T_{obs}}{T} \right\rfloor$ is the number of samples, $c_{i,n}$ the discrete symbol related to the $n$-th sample taken at $t_{i,n} = t_i + nT$, and $\bm{\nu}_{i,n} \sim \mathcal{C}\mathcal{N}_M(\bm{0},\sigma^2 \bm{I}_M)$ the filtered thermal noise, with 
$\sigma^2$ denoting the noise power and  $\bm{I}_M$  the $M \times M$ identity matrix. 
We assume that the symbols are known to the receiver, which is usually obtained by considering the first part of the transmission where a known training sequence is inserted for channel estimation and synchronization purposes \cite{Bazzi_KnownSymbols}.\footnote{A decision-directed approach should also be possible, but is beyond the scope of the present contribution.} Hereafter, we denote with $\bm{Y}_i = [\bm{y}_{i,0} \cdots \bm{y}_{i,N-1}]$ the $M \times N$ matrix containing samples of the $i$-th observation.  It is worth noting that the AOA of the LOS path $\theta^{\text{\tiny LOS}}_i$ directly relates the position of BS $b_i$ to the MS position through
\begin{equation}\label{eq::aoa}
\theta^{\text{\tiny LOS}}_i = \mathrm{atan2} \left(y^{b_i}_{BS} - y(t_i),x^{b_i}_{BS} - x(t_i)\right) 
\end{equation}
where the function $\mathrm{atan2}(y,x)$ is the four-quadrant inverse tangent, and the angle is measured counterclockwise with respect to the $x$-axis, as depicted in Fig.~\ref{fig:ph1}.
{\ed For the MS at a certain time and a given BS, eq. \eqref{eq::aoa} provides a relationship between absolute positions in the reference system $xy$ and corresponding angle, which is thus expressed in the same frame as a number in $(0,2\pi)$ (in Fig.~\ref{fig:ph1} a translation in the current position $\bm{p}(t_i)$ is performed for representation convenience).
It is worth noting that the antenna array can be realigned to such a global reference system by considering, at each time instant $t_i$, a rotation equal to the heading vector obtained from the measured velocity $\bm{v}(t_i)$.
Moreover, angles expressed in the absolute frame $xy$ can be mapped onto the local reference system of the MS, where they are measured  in $(-\pi/2,\pi/2)$ with respect to the normal direction to the array, that is with respect to the heading vector, as shown in Fig. \ref{fig:ph2}.\footnote{\ed We will come back on this issue when discussing the problem of estimating the multipath directions in the derivation of our algorithm.}}

 with the aid of a mobility model. 
For a sufficiently high BSs send rate (e.g., $R_{BS} \geq 10$ Hz), it is reasonable to assume that the time interval between any two consecutive observations is relatively short ($\leq 100$ ms). As a consequence, the (arbitrary) MS trajectory over $[t_0,t_k], k\geq1$ can be approximated by the following kinematic model
\begin{equation}\label{eq::kinmodel}
 	\bm{p}(t_k)=\left[ \begin{array}{ccc}
 		x(t_k) = x_0 + \sum_{i=1}^{k}v_{x}(t_{i-1})(t_i - t_{i-1})  \\
 		y(t_k) = y_0 + \sum_{i=1}^{k}v_{y}(t_{i-1})(t_i - t_{i-1}) \\
 	\end{array} \right]
 \end{equation} 
where a constant velocity vector $\bm{v}(t_i) = \left[ v_x(t_i) \ v_y(t_i)\right]^T$ is considered for $t \in [t_{i}, t_{i+1})$, read from the onboard sensor at time instant $t_i$. 

Differently from most state-of-art solutions {\ed which are based on a single snapshot of the environment}, our approach adds one more dimension to the localization procedure, namely the variation  in time. Thus, although more unknown parameters may need to be estimated, each collected $\bm{Y}_i$ brings a new position-related information that can help the MS to reconstruct its most probable trajectory over time. 




\section{\ed Derivation of the Pseudo Maximum Likelihood algorithm and its competitors}{\label{sec::DPEderiv}}

In this section, we propose a novel pseudo ML algorithm for the system model presented in Sec. \ref{sec:sysmodel}. Let $\mathcal{Y} = \left\{ \bm{Y}_1,\ldots,\bm{Y}_K\right\}$ denote the set of observations available up to the current time instant $t_K$. Assuming that the velocities of the MS are known --- estimates from the onboard
sensors will be used in practice, so we will include velocity errors in the simulations of Sec. \ref{sec:simulation} --- the localization problem reduces to the estimation of the MS (initial) position $\bm{p}_0$, but embedded in a problem with many nuisance parameters due to  multipath propagation.

\subsection{Pseudo ML Position Estimation}
We observe that each sample vector $\bm{y}_{i,n}, 1 \leq i \leq K, 0 \leq n \leq N-1$,
is statistically characterized as
\begin{equation}\label{eq::charsample}
\bm{y}_{i,n} \sim \mathcal{C}\mathcal{N}_M\left( \gamma_i (\bm{x}^{\text{\tiny LOS}}_i +  \bm{x}^{\text{\tiny NLOS}}_i)c_{i,n}, \sigma^2\bm{I}_M\right) 
\end{equation}
where all parameters are treated as deterministic unknowns, except the symbols $c_{i,n}$s, which we recall are assumed known at the receiver. More precisely, the whole set of unknowns includes $\bm{p}_0$, which represents the  parameter of interest, and $\bm{\psi} = (\sigma^2, \bm{\gamma}, \bm{\xi})$ which denotes the vector of nuisance parameters, with $\bm{\gamma}=[\gamma_1\cdots \gamma_K]^T$, and $\bm{\xi}=[\bm{\beta}^T\ \bm{\theta}^T]^T$, with  $\bm{\beta}^T = [\beta^{1}_1\cdots\beta^{D_1}_1 \cdots \beta^{1}_K\cdots\beta^{D_K}_K]$, and $\bm{\theta}^T = [\theta^{1}_1\cdots\theta^{D_1}_1 \cdots \theta^{1}_K \cdots\theta^{D_K}_K ]$.
The  ML direct position estimator is then given by
\begin{equation}\label{eq::DPEML}
\hat{\bm{p}}_0 = \argmax_{\tilde{\bm{p}}_0} \left[\max_{\tilde{\bm{\psi}}} L(\tilde{\bm{p}}_0,\tilde{\bm{\psi}})\right]
\end{equation}
 where $L(\tilde{\bm{p}}_0,\tilde{\bm{\psi}}) \eqdef \log(f(\mathcal{Y}|\tilde{\bm{p}}_0,\tilde{\bm{\psi}}))$ and $f(\punto) $ denotes the probability density function of the observations $\mathcal{Y}$ given $\tilde{\bm{\psi}}$ and an initial position $\tilde{\bm{p}}_0 = \left[ \tilde{x}_0 \ \tilde{y}_0\right]^T$. From \eqref{eq::charsample} it follows that
\begin{align}\label{eq::loglike_1}
L(\tilde{\bm{p}}_0,\tilde{\bm{\psi}}) &= -\left[ MKN\log(\pi \tilde{\sigma}^2) \right. \nonumber\\
& \left. + \frac{1}{\tilde{\sigma}^2}\sum_{i=1}^K\sum_{n=0}^{N-1} \|\bm{y}_{i,n} - \tilde{\gamma}_i(\tilde{\bm{x}}^{\text{\tiny LOS}}_i + \tilde{\bm{x}}^{\text{\tiny NLOS}}_i)c_{i,n}\|^2\right]
\end{align}
where $\| \punto \|$ denotes the vector norm.
 It is worth noting that for a given position trial $\tilde{\bm{p}}_0$, the AOAs of the LOS paths $\{\tilde{\theta}^{\text{\tiny LOS}}_i\}_{i=1}^K$ are readily determined from the computation of $\{\tilde{\bm{p}}(t_i)\}_{i=1}^K$ through \eqref{eq::kinmodel}, followed by the application of the geometric model in \eqref{eq::aoa} {\ed and a proper rotation to map the angle onto the local reference system of the MS in $(-\pi/2,\pi/2)$}. On the other hand, relating in general the nuisance parameters $\tilde{\bm{\psi}}$ to BSs and MS positions  seems not possible. 

We start the resolution of the ML  problem by maximizing with respect to $\tilde{\sigma}^2$. A simple computation shows that
\begin{equation}
\hat{\sigma}^2 = \frac{1}{MKN}\sum_{i=1}^K \sum_{n=0}^{N-1}\|\bm{y}_{i,n} - \tilde{\gamma}_i(\tilde{\bm{x}}^{\text{\tiny LOS}}_i + \tilde{\bm{x}}^{\text{\tiny NLOS}}_i)c_{i,n}\|^2.
\end{equation}
Substituting this value back in \eqref{eq::loglike_1}, neglecting unnecessary constant terms, and considering a monotonic transformation of the log-likelihood function, we obtain the equivalent function
\begin{equation}\label{eq::loglike_2}
\ell(\tilde{\bm{p}}_0,\tilde{\bm{\gamma}},\tilde{\bm{\xi}}) = \sum_{i=1}^K\sum_{n=0}^{N-1} \|\bm{y}_{i,n} - \tilde{\gamma}_i(\tilde{\bm{x}}^{\text{\tiny LOS}}_i + \tilde{\bm{x}}^{\text{\tiny NLOS}}_i)c_{i,n}\|^2
\end{equation}
 and the ML direct position estimator reduces to
\begin{equation}\label{eq::DPEML_2}
\hat{\bm{p}}_0 = \argmin_{\tilde{\bm{p}}_0} \left[\min_{\tilde{\bm{\gamma}},\tilde{\bm{\xi}}} \ell(\tilde{\bm{p}}_0,\tilde{\bm{\gamma}},\tilde{\bm{\xi}})\right].
\end{equation}
Clearly,  minimization of \eqref{eq::loglike_2} with respect to a specific $\tilde{\gamma}_i \in \mathbb{C}$ is equivalent to  minimization of the term $\sum_{n=0}^{N-1}\|\bm{y}_{i,n} - \tilde{\gamma}_i(\tilde{\bm{x}}^{\text{\tiny LOS}}_i + \tilde{\bm{x}}^{\text{\tiny NLOS}}_i)c_{i,n}\|^2$, which yields
\begin{equation}\label{eq::mlgamma}
\hat{\gamma}_i = \frac{ \tilde{\bm{x}}_i^H \bar{\bm{y}}_i}{\|\tilde{\bm{x}}_i\|^2 \bar{\bar c}_i} \qquad i=1,\ldots,K
\end{equation}
where $^H$ is the Hermitian operator, $\tilde{\bm{x}}_i \eqdef \tilde{\bm{x}}^{\text{\tiny LOS}}_i + \tilde{\bm{x}}^{\text{\tiny NLOS}}_i$, $\bar{\bm{y}}_i \eqdef \sum_{n=0}^{N-1} \bm{y}_{i,n}c_{i,n}^*$, and $\bar{\bar c}_i \eqdef \sum_{n=0}^{N-1}|c_{i,n}|^2$.
Substituting back in \eqref{eq::loglike_2} leads to
\begin{equation}\label{eq::loglike_3}
\ell'(\tilde{\bm{p}}_0,\tilde{\bm{\xi}}) = \sum_{i=1}^K \left(\bar{\bar y}_i - \frac{\|\bm{P}_{\tilde{\bm{x}}_i}\bar{\bm{y}}_i \|^2}{\bar{\bar c}_i}\right)
\end{equation}
with $\bar{\bar y}_i \eqdef \sum_{n=0}^{N-1}\|\bm{y}_{i,n}\|^2$ while $\bm{P}_{\tilde{\bm{x}}_i} = \tilde{\bm{x}}_i\tilde{\bm{x}}^{H}_i/\|\tilde{\bm{x}_i}\|^2$ denotes the projector onto the one-dimensional space generated by $\tilde{\bm{x}}_i$. Interestingly, \eqref{eq::loglike_3} is parameterized by the MS initial position $\bm{p}_0$ and the nuisance parameters related to the  NLOS paths, that is $\bm{\xi}$. Keeping in mind that the value of $\tilde{\bm{x}}^{\text{\tiny LOS}}_i$ is uniquely determined for each position hypothesis $\tilde{\bm{p}}_0$, the computation of $\bm{P}_{\tilde{\bm{x}}_i}$ requires the estimation of both NLOS amplitudes ($\beta_i^{m}$s) and AOAs ($\theta_i^{m}$s). However, a direct estimation of the multipath environment from \eqref{eq::loglike_3} is not possible in this case since, in contrast to static scenarios, the number of unknown NLOS parameters significantly increases with the observation size. 
To overcome this drawback, {\ed we resort to the pseudo ML methodology \cite{Gong} illustrated in Sec. \ref{sec:introduction} and} propose to reconstruct an estimate of $\bm{P}_{\tilde{\bm{x}}_i}$ using the following alternative approach:
\begin{enumerate}
\item estimation of the AOAs (both LOS and NLOS) for each observation $\bm{Y}_i \in \mathcal{Y}$ using the smooth-MUSIC algorithm for coherent environment;
\item adoption of an adaptive beamforming strategy that exploits a directional response of the array towards the estimated AOAs to estimate the related amplitudes;
{\ed \item  association of the most likely direction (among the ones estimated in step 1) to the LOS, given a trial initial position $\tilde{\bm{p}}_0$.}
\end{enumerate}

In the following we detail such a procedure.
Spatial smoothing (SS) is an effective way to  decorrelate signals for some array structures  \cite{ShanKailath2}. In particular, the $M$-element ULA is divided into $S$ virtual overlapped subarrays with each subarray composed of $P < M$ sensors and shifted by one with respect to the previous one.\footnote{Denoting with $p$ the first sensor of $j$-th subarray, the first sensor belonging to the $(j+1)$-th subarray is at position $p+1$.} As a result, the full array is divided into $S = M - P + 1$ subarrays. Each set of subarray data is denoted by $\bm{y}_{i,n}^{(j)}, j=1,\ldots,S$, and contains the $P$ components of $\bm{y}_{i,n}$ from $j$ to $j+P-1$, respectively.
The forward-only (FO) matrix is then obtained by using the averaged sample covariance matrix $\hat{\bm{R}}^{\text{\scriptsize F}}_{Y_iY_i} \in \mathbb{C}^{P \times P}$, which is defined as
\begin{equation}\label{eq::fomatrix}
\hat{\bm{R}}^{\text{\scriptsize F}}_{Y_iY_i} = \frac{1}{S} \sum_{j=1}^{S} \hat{\bm{R}}^{(j)}_{Y_iY_i}
\end{equation}
with $\hat{\bm{R}}^{(j)}_{Y_iY_i} = (1/N)\bm{Y}^{(j)}_i\bm{Y}^{(j)H}_i$ denoting the $j$-th subarray sample covariance matrix and $\bm{Y}^{(j)}_i \eqdef [\bm{y}^{(j)}_{i,0} \cdots \bm{y}^{(j)}_{i,N-1}]$. 
Better, a forward-backward spatial smoothing (FBSS) can be employed to decorrelate the received signal in a stronger way; after that, a MUSIC approach can be used, referred to as smooth-MUSIC in this case \cite{twodecades}. More precisely, let $\bm{J}$ be an exchange matrix, whose elements are zero except for ones on the antidiagonal. By exploiting the translational invariance property of $\bm{a}(\theta_i)$, i.e., $\bm{J}\bm{a}^*(\theta_i)=  e^{-j(M-1)\omega d\sin\theta_i}\bm{a}(\theta_i)$, the following forward-backward sample covariance matrix can be used in place of \eqref{eq::fomatrix}
\begin{equation}\label{eq::R_FBSS}
\hat{\bm{R}}^{\text{\scriptsize FB}}_{Y_iY_i} = \frac{\hat{\bm{R}}^{\text{\scriptsize F}}_{Y_iY_i} + \bm{J}(\hat{\bm{R}}^{\text{\scriptsize F}}_{Y_iY_i})^* \bm{J}}{2}.
\end{equation}
Considering without loss of generality the first subarray as the reference subarray, we denote its steering vector as $\bm{a}^{(1)}(\theta_i) = \left[1 \ e^{j\omega d\sin\theta_i} \cdots \ e^{j(P-1)\omega d\sin\theta_i}\right]^T$ and compute the smooth-MUSIC algorithm on the reduced-size vector.

For each observation $i$, let us denote by $\hat{\bm{\theta}}_i=[\hat{\theta}_i^1\cdots \hat{\theta}_i^{D_{\text{max}}+1}]^T$ the estimates obtained by the described procedure, i.e., $\hat{\theta}_i^s$, $s=1,\ldots,D_{\text{max}}+1$ are the directions corresponding to the peaks of the smooth-MUSIC pseudo-spectrum. Clearly, the number $D_{\text{max}}+1$ of estimated components can be different from the actual $D_i+1$ (LOS+NLOS); nonetheless, one can expect that if there are spurious directions due to  variations in the pseudo-spectrum, insignificant amplitudes  will be obtained when searching through such directions; {\ed similarly, if some directions are  missed due to a $D_i > D_{\text{max}}$, they will be reasonably the least significant in amplitude hence their residual effect should be  limited (as confirmed by the simulations shown later).} For all such $D_{\text{max}}+1$ components, the complex amplitudes  are estimated as the output of a FBSS Capon beamformer; notice that it will estimate  the product of three terms: $\beta_i^s$ --- with $s\in\{1,\ldots,D_{\text{max}}+1\}$, ideally {\ed equal to one for the LOS component or close to one of the $\beta_i^m$ for the NLOS paths} --- times $\gamma_i$ times the array gain in the look direction. We denote  by $\hat{\alpha}_i^s$  the overall estimated amplitudes:
\begin{equation}\label{eq::alphaesti}
\hat{\alpha}_i^s = \bm{w}_{\text{\scriptsize FB}}^H(\hat{\theta}_i^s)\bm{y}_{i,n}^{(1)}
\end{equation}
with the optimum weight vector $\bm{w}_{\text{\scriptsize FB}}\in \mathbb{C}^{P \times 1}$ given by \cite{vantrees}
\begin{equation}\label{eq::FBSSweight_first}
\bm{w}_{\text{\scriptsize FB}}(\hat{\theta}_i^s)= \frac{(\hat{\bm{R}}^{\text{\scriptsize FB}}_{Y_iY_i})^{-1}\bm{a}^{(1)}(\hat{\theta}_i^s)}{\bm{a}^{(1)H}(\hat{\theta}_i^s)(\hat{\bm{R}}^{\text{\scriptsize FB}}_{Y_iY_i})^{-1}\bm{a}^{(1)}(\hat{\theta}_i^s)}.
\end{equation}

The vector of estimated amplitudes is denoted by $\hat{\bm{\alpha}}_i$. To  reconstruct a meaningful estimate of the projection matrix for the final step of the ML estimation procedure, we need to consider that also $\beta_0$ (the component related to the  LOS), although theoretically equal to 1, is estimated this way (it is one of the $D_{\text{max}}+1$ directions). Thus, in the estimated projector all components (LOS+NLOS) share the same estimate of $\gamma_i$, which therefore becomes a constant that cancels out in the  normalization intrinsic in $\bm{P}_{\tilde{\bm{x}}_i} = \tilde{\bm{x}}_i\tilde{\bm{x}}^{H}_i/\|\tilde{\bm{x}_i}\|^2$. Thus, up to (minor) errors due to the array gain not being perfectly equal to 1 in the look direction --- to the extent of the angle estimation errors from the smooth-MUSIC --- the projection matrix can be reconstructed.
To this aim, the main problem remains  the identification of the LOS, i.e., the association of one of the estimated directions $\hat{\theta}_i^s$ to the direct path, in order to separate the complementary NLOS components involved in $\bm{P}_{\tilde{\bm{x}}_i}$ from the LOS component that is integrated over the trajectory for $i=1,\ldots,K$. We proceed as follows.
\begin{itemize}
\item For each trial position $\tilde{\bm{p}}_0$ in a  grid, and based on the trajectory resulting from the velocity measurements up to time $t_i$, we can reconstruct the trial LOS angle $\tilde{\theta}^{\text{\tiny LOS}}_i$. {\ed Such an angle is mapped to the local reference system of the array, i.e., in the interval $(-\pi/2,\pi/2)$ with respect to the heading vector, so spanning the front half-plane of the mobile node (first and second quadrant in its local frame). Such a map is non-invertible due to the symmetry of revolution of the ULA, therefore directions from the rear half-plane (third and fourth quadrant in the local frame) will be folded to the corresponding symmetric (front) angles as ``seen'' by the array.}
\item This direction is compared against the estimated AOAs at $i$: if a $\hat{\theta}_i^\star$ in $\hat{\bm{\theta}}_i$ is found such that its distance to $\tilde{\theta}^{\text{\tiny LOS}}_i$  is less than a tolerance $\delta$ (namely, a few degrees), then $\hat{\theta}_i^\star$ is associated to the LOS\footnote{\ed If multiple angles are found that are close to $\tilde{\theta}^{\text{\tiny LOS}}_i$ by less than $\delta$, clearly the closest among them is associated to the LOS.}; as a consequence, its estimated amplitude $\hat{\alpha}_i^\star$  (taken from $\hat{\bm{\alpha}}_i$) is used to compute  an estimate of the LOS signal $\hat{\tilde{\bm{x}}}^{\text{\tiny LOS}}_i = \hat{\alpha}_i^\star\bm{a}(\tilde{\theta}^{\text{\tiny LOS}}_i)$ --- notice that $\tilde{\theta}^{\text{\tiny LOS}}_i$ is used in the reconstruction, not $\hat{\theta}_i^\star$. 
\item Likewise, we compute an estimate of the NLOS signal as $\hat{\tilde{\bm{x}}}^{\text{\tiny NLOS}}_i = \sum_{j}\hat{\alpha}^{\text{\tiny NLOS}}_j\bm{a}(\hat{\theta}^{\text{\tiny NLOS}}_j)$ with $\hat{\alpha}^{\text{\tiny NLOS}}_j \in \hat{\bm{\alpha}}_i \setminus \{\hat{\alpha}_i^\star\}$ and $\hat{\theta}^{\text{\tiny NLOS}}_j \in \hat{\bm{\theta}}_i \setminus \{\hat{\theta}_i^\star\}$. This yields an  estimate of the projector over $\tilde{\bm{x}}_i = \tilde{\bm{x}}^{\text{\tiny LOS}}_i + \tilde{\bm{x}}^{\text{\tiny NLOS}}_i$, i.e., $\hat{\bm{P}}_{\tilde{\bm{x}}_i}$, to be used in the final optimization of the (compressed) pseudo likelihood function, i.e.,
\begin{equation}\label{eq::loglike_4}
\hat{\bm{p}}_0 = \argmax_{\tilde{\bm{p}}_0\in\mathcal{P}}  \sum_{i\in\mathcal{A}(\tilde{\bm{p}}_0)}  \frac{\|\hat{\bm{P}}_{\tilde{\bm{x}}_i}\! (\tilde{\bm{p}}_0)\,\bar{\bm{y}}_i \|^2}{\bar{\bar c}_i}
\end{equation}
where we have remarked the dependency of the projection matrix on $\tilde{\bm{p}}_0$.
\end{itemize}

\begin{figure}
	\centering
	\includegraphics[width=0.6\textwidth]{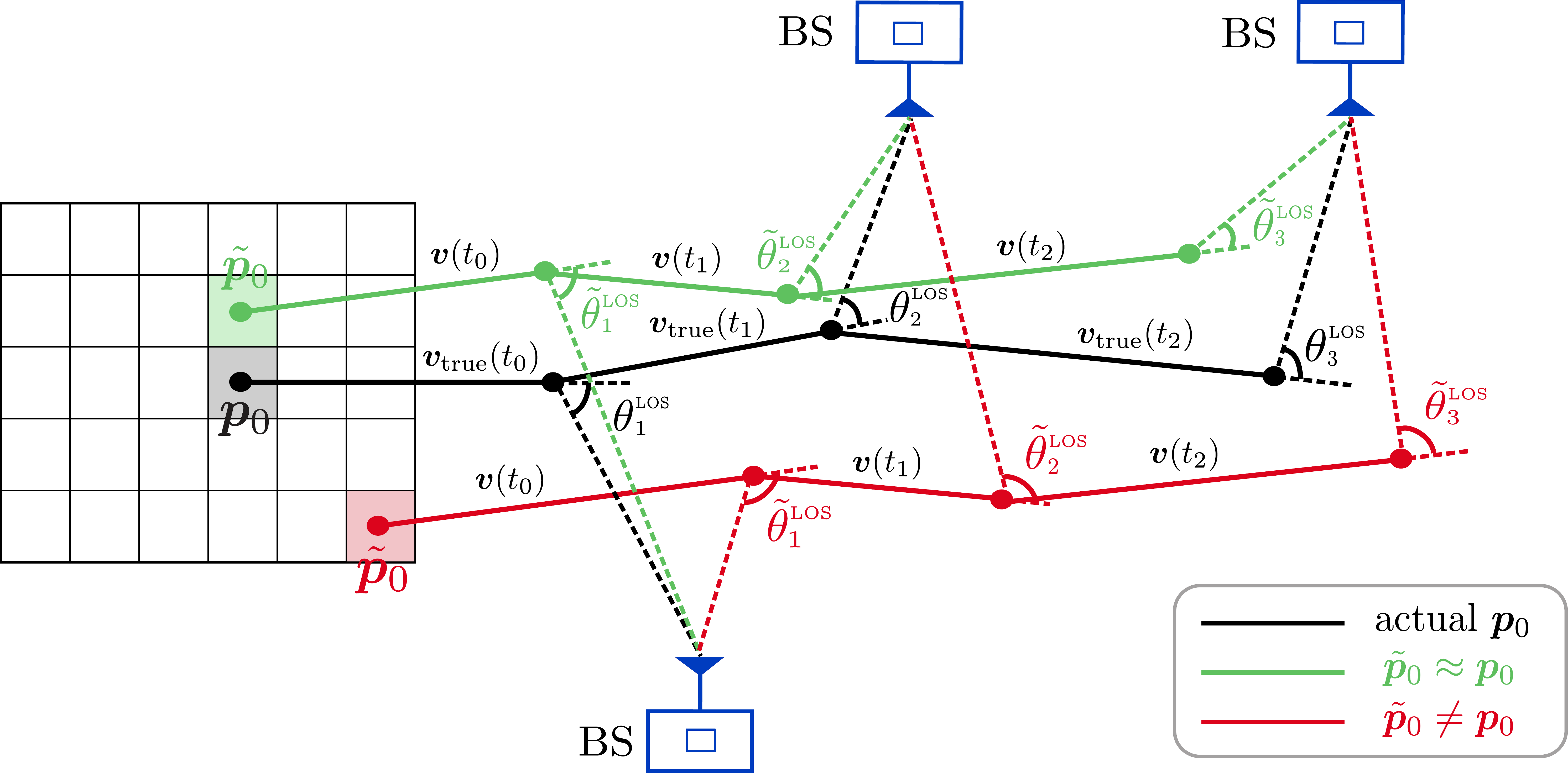}
	\caption{Example of trajectory/angle reconstruction up to $K=3$ for two position trials $\tilde{\bm{p}}_0$ and true trajectory, for $N_{BS}=3$.}
	\label{fig:ph2}
\end{figure}

\begin{figure}
\centering
	\includegraphics[width=0.4\textwidth]{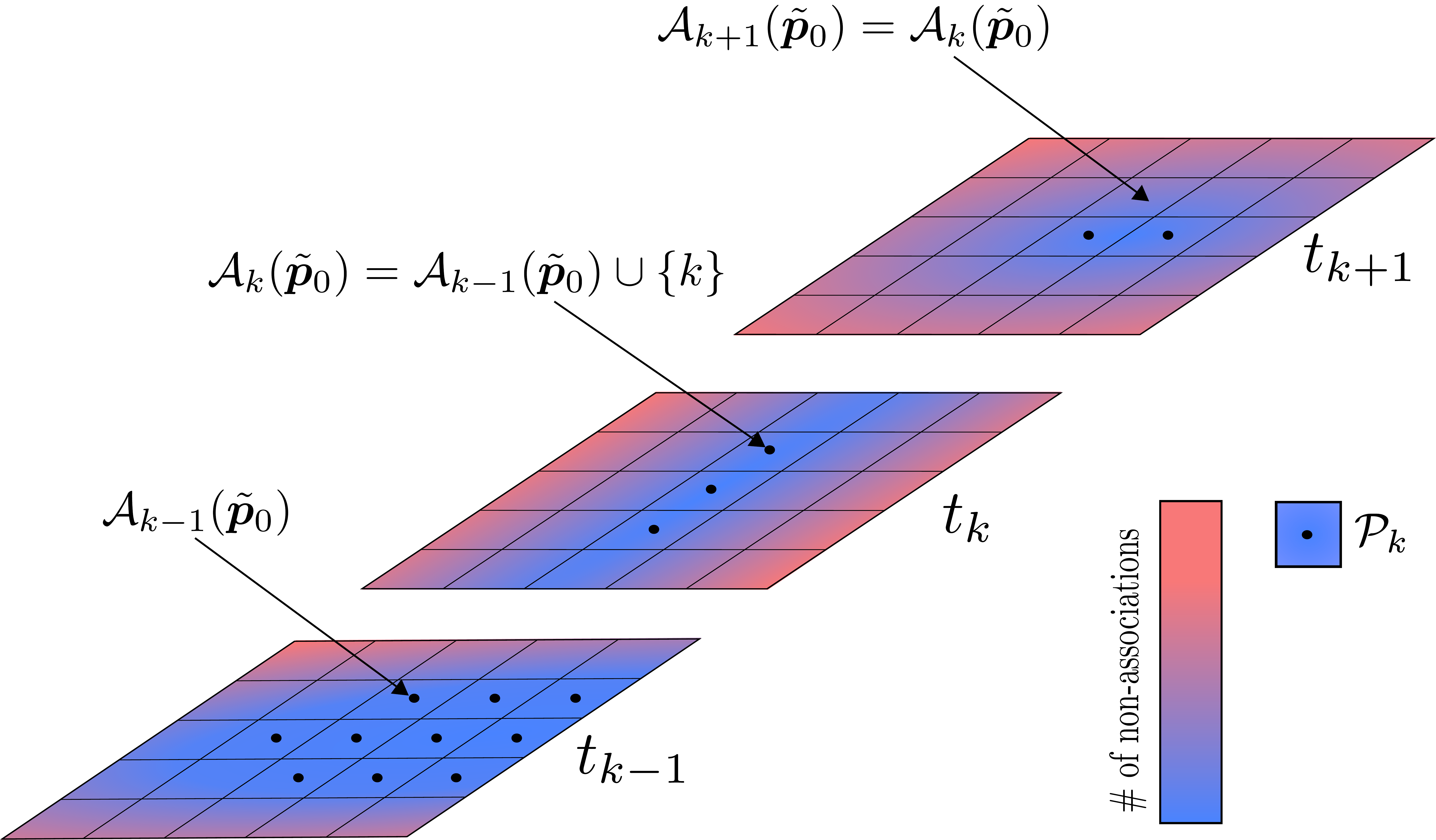}
	\caption{Pictorial representation of a possible evolution over time for the sets $\mathcal{A}_k$ and $\mathcal{P}_k$.}
	\label{fig:ph3}
\end{figure}

{\ed It is worth highlighting that angles cannot be estimated unambiguously through a linear array. However, the association mechanism does not need to know the true angles: it is sufficient that a compatible angle is found in the smooth-MUSIC pseudo-spectrum, irrespective of which of the two possible directions that lead to the same steering vector is the true one. Indeed, in the projection matrix only the steering vector appears, not the true angle, meaning that the ambiguity is not a problem for the computation of the cost function of our algorithm.}

Notice also that in \eqref{eq::loglike_4} the sum is taken  on the subset $\mathcal{A}(\tilde{\bm{p}}_0)$ of indexes for which the LOS association has been performed. In fact, if in $\hat{\bm{\theta}}_i$ there is no estimated direction sufficiently close to the   LOS trajectory under evaluation, obtained from a given trial point $\tilde{\bm{p}}_0$, such $i$-th term is discarded from the cost function. The algorithm will keep state of the number of indexes not associated (i.e., terms discarded in the cost function) for each trial point $\tilde{\bm{p}}_0$ in the grid; once the evaluation of all points is concluded, the maximum will be taken only on the subset $\mathcal{P}$ of  grid points with minimum number of non-associations, since  the  likelihood of LOS association is maximized for points close to the true one. Fig. \ref{fig:ph2} illustrates this idea by showing two examples of reconstructed trajectories based on different $\tilde{\bm{p}}_0$, in comparison with the true trajectory (black curve in the middle). Clearly, all ordered segments of the reconstructed trajectories are parallel to each other since they use the same velocity estimates ($\bm{v}(t_0),\bm{v}(t_1), \ldots$), with some misalignment compared to the true trajectory due to the measurement errors. Thus, it is expected that a $\tilde{\bm{p}}_0$ closer to the true $\bm{p}_0$ will produce a larger value in the cost function. {\ed Notice also that the geometric direction of the LOS may be outside the interval $(-\pi/2,\pi/2)$ in which angles can be estimated by the ULA, but as explained this is handled naturally by the proposed association mechanism.}

To further clarify, Fig. \ref{fig:ph3} depicts an example of possible evolution over time for the set $\mathcal{P}$. The algorithm can be implemented in an on-line fashion since the association decision for past observations does not change over time. In particular, denoting by  $\mathcal{P}_k$ the subset of  grid points with minimum number of non-associations at time~$k$, eq. \eqref{eq::loglike_4} can be rewritten in a recursive form as follows
\begin{equation}\label{eq::loglike_recursive}
\hat{\bm{p}}_0 (k) = \argmax_{\tilde{\bm{p}}_0\in\mathcal{P}_k}  S_{k}(\tilde{\bm{p}}_0)
\end{equation}
where
$$
S_{k}(\tilde{\bm{p}}_0) = S_{k-1}(\tilde{\bm{p}}_0) + \delta_k(\tilde{\bm{p}}_0)   \frac{\|\hat{\bm{P}}_{\tilde{\bm{x}}_k}\bar{\bm{y}}_k \|^2}{\bar{\bar c}_k}
$$
with
$$
S_{k-1}(\tilde{\bm{p}}_0) = \sum_{i\in\mathcal{A}_{k-1}(\tilde{\bm{p}}_0)}  \frac{\|\hat{\bm{P}}_{\tilde{\bm{x}}_i}\bar{\bm{y}}_i \|^2}{\bar{\bar c}_i},
$$
$$
\delta_k(\tilde{\bm{p}}_0)=
\left\{ \begin{array}{ll} 1 & \text{if $k$-th meas. associated to LOS given $\tilde{\bm{p}}_0$} \\ 
0 & \text{otherwise} 
\end{array} \right.
$$
and
$$
\mathcal{A}_{k}(\tilde{\bm{p}}_0) =  \left\{ \begin{array}{ll} \mathcal{A}_{k-1}(\tilde{\bm{p}}_0) \cup \{k\} & \text{if $\delta_k(\tilde{\bm{p}}_0) = 1$} \\ 
\mathcal{A}_{k-1}(\tilde{\bm{p}}_0) & \text{otherwise} 
\end{array} \right. .
$$

As a final remark, we observe that the proposed algorithm intrinsically handles transmissions from multiple BS,  ``interleaved'' in the time index $i$ (according to the arrival time) without any additional complexity. It is only sufficient to consider for each transmission the appropriate BS position.

For the sake of clarity, the algorithm steps are summarized in Algorithm \ref{alg::MLDPE}.
Notice that, despite the algorithmic procedure is articulated in several steps, the computational complexity is limited   thanks to the recursive implementation.\footnote{\ed As mentioned, the proposed pseudo ML algorithm can be in principle extended to the 3D case, in which also elevation angles are considered. In fact, the properties required in the FBSS step to decorrelate the multipath, i.e., translational invariance and Vandermonde's structure, are fulfilled not only by ULAs but also by uniform rectangular arrays (URAs); the proposed pseudo ML approach can be in turn extended to cope with both azimuth and elevation angles, using the well-known two-dimensional variants of the MUSIC algorithm and Capon beamformer, and finally performing the position optimization on a 3D grid instead of a 2D one. The computational complexity of the procedure, of course, would be higher as in any higher-dimensional problem, but no additional theoretical issues arise.}

\begin{algorithm}
\caption{ML-based DPE (online implementation)}\label{alg::MLDPE}
\begin{algorithmic}[1]
\State \textbf{Initialize:} 
\State Set $k = 0$, $\mathcal{P}_0 = \left\{(x,y) \, \text{of a chosen 2D grid} \right\}$
\For{each $\tilde{\bm{p}}_0 \in \mathcal{P}_0$}
\State $\mathcal{A}_0(\tilde{\bm{p}}_0) = \{\emptyset\}$
\State $S_0(\tilde{\bm{p}}_0) = 0$
\State $I(\tilde{\bm{p}}_0) = 0$
\EndFor
\Loop 
\State New observation: $k \gets k + 1$ (process $\bm{Y}_k$ at $t_k$)
\State Compute $\hat{\bm{\theta}}_k$ using the smooth-MUSIC algorithm
\For{each $\hat{\theta}_k^s$ in $\hat{\bm{\theta}}_k$}
        \State Compute beamformer weights $\bm{w}_{\text{\scriptsize FB}}(\hat{\theta}_k^s)$ using \eqref{eq::FBSSweight_first}
        \State Compute $\hat{\alpha}_k^s$ using \eqref{eq::alphaesti}
      \EndFor
      \For{each $\tilde{\bm{p}}_0 \in \mathcal{P}_0$}
      \State Compute $\tilde{\theta}^{\text{\tiny LOS}}_k$ using \eqref{eq::kinmodel} 
      \For{each $\hat{\theta}_k^s$ in $\hat{\bm{\theta}}_k$}
      \State Compute $d^s_k = |\tilde{\theta}^{\text{\tiny LOS}}_k - \hat{\theta}_k^s|$
      \EndFor
      \If{$\displaystyle \min_{s=1,\ldots,q_k} \!\!\!\left\{d_k^s \right\} \leq \delta$}
      \State Compute $\hat{\bm{P}}_{\tilde{\bm{x}}_k}\! (\tilde{\bm{p}}_0)$
      \State Compute $S_{k}(\tilde{\bm{p}}_0) = S_{k-1}(\tilde{\bm{p}}_0) +  \frac{\|\hat{\bm{P}}_{\tilde{\bm{x}}_k}\bar{\bm{y}}_k \|^2}{\bar{\bar c}_k}$
      \State Update $\mathcal{A}_{k}(\tilde{\bm{p}}_0) = \mathcal{A}_{k-1}(\tilde{\bm{p}}_0) \cup \{ k\}$
      \Else
      \State $I(\tilde{\bm{p}}_0) \gets I(\tilde{\bm{p}}_0) + 1$
      \EndIf
      \EndFor
      \State $\mathcal{P}_k = \left\{\tilde{\bm{p}}_0 \in \mathcal{P}_0 \; \text{s.t.} \; I(\tilde{\bm{p}}_0) = \min I\right\}$
      \State Compute $\displaystyle \hat{\bm{p}}_0 (k) = \argmax_{\tilde{\bm{p}}_0\in\mathcal{P}_k}  S_{k}(\tilde{\bm{p}}_0)$
      \State Reconstruct $\hat{\bm{p}}(t_k)$ using \eqref{eq::kinmodel}
      \EndLoop
\end{algorithmic}
\end{algorithm}

{\ed The approach proposed in this section to deal with mobility can be also adopted to extend state-of-the-art algorithms designed for a static scenario, so obtaining natural competitors for the pseudo ML algorithm.
In particular, the techniques proposed in \cite{weiss2,weiss1} and \cite{weiss4,weiss5}
can be modified to cope with the same scenario we are addressing in this paper: i) by considering  downlink reception from one or multiple BSs, and ii) by using onboard velocity estimates to make a direct position estimation based on a batch of signals collected over multiple past time instants, suitably sampled as discussed in Sec. \ref{sec:sysmodel}. The derivations are reported in the next two subsections.
}

\subsection{Max-power DPE}

{\ed 
Starting from \cite{weiss4,weiss5}, the technique proposed therein uses a white Gaussian model coupled with the optimization of a standard MVDR cost function. Since in our case the multipath is also considered, the MVDR would fail due to coherence in the received signal; we therefore introduce spatial smoothing also in the cost function of \cite{weiss4,weiss5}, for a fair comparison.
More in details, in the  adaptation of such work to our framework, a compressed version of the observations $\mathcal{Y}$, namely $\mathcal{Z} = \left\{\bm{z}_1(\theta_1),\dots,\bm{z}_K(\theta_K)\right\}$ with $\bm{z}_i(\theta_i) = (\bm{w}_{\text{\scriptsize FB}}^H(\theta_i)\bm{Y}^{(1)}_i)^T$is used}, which are parameterized as function of the look angles $\{\theta_1,\ldots,\theta_K\}$ (we stress the dependencies on $\theta_i$) and with the optimum weight vector $\bm{w}_{\text{\scriptsize FB}}\in \mathbb{C}^{P \times 1}$ given by
\begin{equation}\label{eq::FBSSweight}
\bm{w}_{\text{\scriptsize FB}}(\theta_i)= \frac{(\hat{\bm{R}}^{\text{\scriptsize FB}}_{Y_iY_i})^{-1}\bm{a}^{(1)}(\theta_i)}{\bm{a}^{(1)H}(\theta_i)(\hat{\bm{R}}^{\text{\scriptsize FB}}_{Y_iY_i})^{-1}\bm{a}^{(1)}(\theta_i)}.
\end{equation}
For a given position trial $\tilde{\bm{p}}_0$, each look angle $\tilde{\theta}_i$ is determined by computing $\tilde{\bm{p}}(t_i)$ through \eqref{eq::kinmodel}. 
This leads to
\begin{equation}
\hat{\bm{p}}_0 = \argmax_{\tilde{\bm{p}}_0} \sum_{i=1}^K \| \bm{z}_i(\tilde{\theta}_i) \|^2.
\end{equation}
Such a DPE algorithm is completely different from the pseudo ML algorithm, and does not exploit the knowledge of the  symbols $\bm{c}_{i,n}$.
Intuitively, it  aims at measuring the amount of energy collected over time for each trial position $\tilde{\bm{p}}_0$. In doing so, when $\tilde{\bm{p}}_0 \approx \bm{p}_0$, the look directions would be close to the actual $\{\theta^{\text{\tiny LOS}}_1,\ldots,\theta^{\text{\tiny LOS}}_K\}$ and the cumulative energy will take into account the contributions of LOS paths, which contain considerably higher power than that of all the NLOS components. However, since only a finite number of noisy samples is available, we expect that the estimated energy may exhibit significant deviations from its actual value, especially when a high number of NLOS components is present.

For the sake of clarity, the steps of this algorithm, labeled ``Max-power DPE'' are summarized in Algorithm \ref{alg::MP}.
Notice that the computational complexity is  lower than Algorithm \ref{alg::MLDPE}; as will be shown in the analysis below, this is in trade-off with the localization performance, especially under severe multipath conditions.

\begin{algorithm}
\caption{Max-power DPE}\label{alg::MP}
\begin{algorithmic}[1]
\State \textbf{Initialize:} 
\State Set $k = 0$, $\mathcal{P}_0 = \left\{(x,y) \, \text{of a chosen 2D grid} \right\}$
\For{each $\tilde{\bm{p}}_0 \in \mathcal{P}_0$}
\State $S_0(\tilde{\bm{p}}_0) = 0$
\EndFor
\Loop 
\State New observation: $k \gets k + 1$ (process $\bm{Y}_k$ at $t_k$)
      \State Compute $\tilde{\theta}_k$ using \eqref{eq::kinmodel} 
\State Compute beamformer weights $\bm{w}_{\text{\scriptsize FB}}(\tilde{\theta}_k)$ using \eqref{eq::FBSSweight}
\State Compute $\bm{z}_k(\tilde{\theta}_k) = (\bm{w}_{\text{\scriptsize FB}}^H(\tilde{\theta}_k)\bm{Y}^{(1)}_k)^T$
     \For{each $\tilde{\bm{p}}_0 \in \mathcal{P}_0$}
      \State Compute $S_{k}(\tilde{\bm{p}}_0) = S_{k-1}(\tilde{\bm{p}}_0) +   \| \bm{z}_k(\tilde{\theta}_k) \|^2$
      \EndFor
      \State Compute $\displaystyle \hat{\bm{p}}_0 (k) = \argmax_{\tilde{\bm{p}}_0\in\mathcal{P}_0}  S_{k}(\tilde{\bm{p}}_0)$
      \State Reconstruct $\hat{\bm{p}}(t_k)$ using \eqref{eq::kinmodel}
      \EndLoop
\end{algorithmic}
\end{algorithm}

\subsection{Single-path ML DPE}

In this section we derive a single-path (SP) ML DPE algorithm that ignores the NLOS components. Notice that this is tantamount to considering that all multipath effects are modeled as additive white Gaussian noise, as done in \cite{weiss2,weiss1}.
However, as mentioned, we cannot directly take \cite{weiss2,weiss1} as competitors since they are for stationary, not mobile nodes, {\ed and consider a different problem setup. 
In the following we provide the necessary adaptation to make them consistent with our framework.

Starting from the multipath-free white-noise model assumption in \cite{weiss2,weiss1}, in our case we have that}
\begin{equation}
\bm{y}_{i,n} \sim \mathcal{C}\mathcal{N}_M\left( \gamma_i\bm{x}^{\text{\tiny LOS}}_ic_{i,n}, \sigma^2_{\text{\tiny SP}}\bm{I}_M\right) \label{eq:SPmodel}
\end{equation}
where  $\sigma^2_{\text{\tiny SP}}$ denotes the ultimate variance accounting for both thermal noise and NLOS contributions, and $\bm{x}^{\text{\tiny LOS}}_i = \bm{a}\left(\theta^{\text{\tiny LOS}}_i\right)$ as usual.  
The  log-likelihood function is expressed as
\begin{align}\label{eq::MLloglike}
L(\tilde{\bm{p}}_0,\tilde{\sigma^2_{\text{\tiny SP}}},\tilde{\bm{\gamma}}) = -&\left[ MKN\log(\pi \tilde{\sigma}^2_{\text{\tiny SP}}) \right. \nonumber\\
& \left. + \frac{1}{\tilde{\sigma}^2_{\text{\tiny SP}}}\sum_{i=1}^K \sum_{n=0}^{N-1} \|\bm{y}_{i,n} - \tilde{\gamma}_i\bm{a}(\tilde{\theta}_i) c_{i,n}\|^2 \right]
\end{align}
where, again, for a given position trial $\tilde{\bm{p}}_0$ the resulting trial LOS directions $\tilde{\theta}_i, i=1,\ldots,K$, are obtained from the application of \eqref{eq::kinmodel} and \eqref{eq::aoa}. The maximum of \eqref{eq::MLloglike} with respect to $\tilde{\sigma}^2_{\text{\tiny SP}}$ is given by
\begin{equation}
\hat{\sigma}^2_{\text{\tiny SP}} = \frac{1}{MKN}\sum_{i=1}^K \sum_{n=0}^{N-1}\|\bm{y}_{i,n} - \tilde{\gamma}_i\bm{a}(\tilde{\theta}_i) c_{i,n}\|^2. 
\end{equation}
Substituting this expression back in \eqref{eq::MLloglike}, neglecting unnecessary constant terms, and considering a monotonic transformation of the log-likelihood, we obtain
\begin{equation}\label{eq::MLloglike2}
\ell_{\text{\tiny SP}}\left(\tilde{\bm{p}}_0,\tilde{\bm{\gamma}}\right) \!=\!  \sum_{i=1}^K \sum_{n=0}^{N-1}\|\bm{y}_{i,n} - \tilde{\gamma}_i\bm{a}(\tilde{\theta}_i) c_{i,n}\|^2\!\!.
\end{equation}
It is a simple matter to show that  maximization of \eqref{eq::MLloglike2} with respect to a specific $\tilde{\gamma}_i \in \mathbb{C}$ is solved by
\begin{equation}\label{eq::gammahat}
\hat{\gamma}_i = \frac{\sum_{n=0}^{N-1} |\bm{a}^H(\tilde{\theta}_i) \bm{y}_{i,n}c_{i,n}^* |^2}{\sum_{n=0}^{N-1} \|\bm{a}(\tilde{\theta}_i) c_{i,n} \|^2} \quad i=1,\ldots,K.
\end{equation}
Substituting  these maximizing values back in \eqref{eq::MLloglike2}, the final ML DPE is obtained as
\begin{equation}
\hat{\bm{p}}_0 = \argmin_{\tilde{\bm{p}}_0} \sum_{i=1}^K \sum_{n=0}^{N-1}\|\bm{y}_{i,n} - \hat{\gamma}_i\bm{a}(\tilde{\theta}_i) c_{i,n}\|^2.
\end{equation}
{\ed 
For completeness, we summarize the steps of the SP estimator in Algorithm \ref{alg::SP}. 
As it can be noticed, this approach does not exploit any beamforming procedure to cope with multipath; therefore, it is reasonable to expect that its performance are inferior. Moreover, as  discussed, it shares some similarity with the approaches proposed in \cite{weiss2,weiss1} only in the cost function used to perform the ultimate position estimation.
\begin{algorithm}\ed 
\caption{Single-path DPE}\label{alg::SP}
\begin{algorithmic}[1]
\State \textbf{Initialize:} 
\State Set $k = 0$, $\mathcal{P}_0 = \left\{(x,y) \, \text{of a chosen 2D grid} \right\}$
\For{each $\tilde{\bm{p}}_0 \in \mathcal{P}_0$}
\State $S_0(\tilde{\bm{p}}_0) = 0$
\EndFor
\Loop 
\State New observation: $k \gets k + 1$ (process $\bm{Y}_k$ at $t_k$)
      \State Compute $\tilde{\theta}_k$ using  \eqref{eq::kinmodel} 
\State Compute $\hat{\gamma}_k$ using \eqref{eq::gammahat}
\State Compute $r_k(\tilde{\theta}_k) \eqdef \sum_{n=0}^{N-1}\|\bm{y}_{k,n} - \hat{\gamma}_k\bm{a}(\tilde{\theta}_k) c_{k,n}\|^2$
     \For{each $\tilde{\bm{p}}_0 \in \mathcal{P}_0$}
      \State Compute $S_{k}(\tilde{\bm{p}}_0) = S_{k-1}(\tilde{\bm{p}}_0) +  r_k(\tilde{\theta}_k)$
      \EndFor
      \State Compute $\displaystyle \hat{\bm{p}}_0 (k) = \argmax_{\tilde{\bm{p}}_0\in\mathcal{P}_0}  S_{k}(\tilde{\bm{p}}_0)$
      \State Reconstruct $\hat{\bm{p}}(t_k)$ using \eqref{eq::kinmodel}
      \EndLoop
\end{algorithmic}
\end{algorithm}


%


In the following, we compare the proposed pseudo ML approach (Algorithm \ref{alg::MLDPE})
against the Max-power DPE (Algorithm \ref{alg::MP}) and single-path DPE (Algorithm \ref{alg::SP}). Again, we remark that the latter are representative of the most natural competitors available in the literature, but  differ from \cite{weiss2,weiss1} and \cite{weiss4,weiss5}, respectively, 
which are designed for a different localization problem setup.
}  

\section{Simulation model and results}\label{sec:simulation}
In this section, we analyze the performance of the  algorithms by means of simulations. We consider a MS equipped with a $M = 64$ element ULA, running the localization algorithms, and different test scenarios with one or more BSs that broadcast signals with a rate $R_{BS} = 10$ Hz. To simulate a realistic environment, we model several phenomena and non-idealities that can be found in real contexts. It is worth remarking, thus, that the performance assessment is carried out in a simulation environment that is not matched to the design assumptions of the proposed algorithms. We consider the root mean squared error (RMSE) as performance metric, estimated based on 200 Monte Carlo trials.

\subsection{Simulation model}
In the following, we give a detailed description of the models adopted for the simulation. 

\subsubsection{Mobility model} 
{\ed We assume that the MS proceeds along a non-linear trajectory starting from the initial position $\bm{p}_0 = [13 \ 7]^T$ [m], with constant transversal acceleration of about 0.025 [m/s${}^2$] aimed at emulating a small turn on the right. The overall velocity profile} is characterized by an accelerated motion for one third of the path (from 25 to 50 km/h in modulus), followed by a constant velocity motion at modulus 50 km/h for the second third, and ending with a deceleration until reaching the initial speed of 25 km/h, 
resulting in a total path of 8 seconds.  This pattern simulates some possible kinematic variations typical of a mobile scenario. MS velocity measurement errors are modeled as independent Gaussian variables with zero mean and standard deviations equal to $10\%$ of the true value of the velocities. 

\subsubsection{Channel model}
We assume a carrier frequency $f_c = 5.9$  GHz and a transmit power $P_{T,dB} = 18$ dBm.\footnote{Such values are typical of mobile communications based e.g. on the IEEE 802.11p standard, and are compatible with the recently-proposed 5G specifications.} The wireless propagation is modeled according to \cite{rappaport}; in particular, the path loss at distance $d$ from the transmitter is obtained by the well-known formula
\begin{equation}
L_{PL,dB} = 10\eta \log_{10}\frac{d}{d_0}
\end{equation}
with path loss exponent  $\eta = 4$ and $d_0 = 1$ m the reference distance. 
According to the experimental campaign conducted in \cite{bernado}, we set the channel parameters $B_c = 250$ kHz and $B_D =  512$ Hz, which represent a harsh multipath environment.

Following the Clarke's model \cite{clarke} for a MS moving in rich multipath environments, we assume that the NLOS contributions can arrive at the receiver from all directions, uniformly distributed in the space, i.e., $\theta^m_i \sim \mathcal{U}(0,2\pi)$. Each complex multipath coefficient can be expressed in terms of $\beta^m_i = a^m_i\e^{j\varphi^m_i}$ with $a^m_i$ the amplitude of the $m$-th NLOS component and $\varphi^m_i = 2\pi f_c \tau^m_i$ the phase shift related to the time delay $\tau^m_i$, respectively. {\ed Each delay $\tau^m_i$ is generated according to the corresponding (random) phase $\varphi^m_i$. In particular, to account also for larger delays (related to longer NLOS paths), we adopt as actual $\tau^m_i$ the sum of the value obtained by inverting the phase $\varphi^m_i \in [0,2\pi]$ plus an increment equal to $\frac{\zeta}{f_c}$, with $\zeta$ a random integer uniformly distributed between 0 and 4}. Following \cite{rappaport}, we model $\varphi^m_i \sim \mathcal{U}(0,2\pi)$, while the multipath amplitude $a^m_i$ is chosen according to a deterministic power delay profile (PDP) $P(\tau^m_i)$, which accounts for the propagation loss as function of the time delay $\tau^m_i$. We have chosen $P(\tau^m_i)$ as an exponential decaying function of $\tau^m_i$, i.e.
\begin{equation}
P(\tau^m_i) = \e^{-\tau^m_i/\sigma_{\tau}}
\end{equation}
with $\sigma_{\tau} = 677$ ns denoting the channel RMS delay spread, set according to \cite{bernado}.

To take into account  LOS obstructions, 10\% of the links are randomly assigned to the NLOS class, while we recall that all algorithms assume there is always a direct path. {\ed As we have seen in Sec. \ref{sec:sysmodel}, the number of multipath components $D_i$ is typically unknown, hence the proposed algorithm will assume a model order equal to $D_{\text{max}}$ while the actual $D_i$ is generally different.  In doing so, we will investigate the sensibility of the proposed algorithms to a misknowledge of the multipath environment, as typical in real scenarios.} As for the symbols, we assume a QPSK constellation for generating the random sequence $c_{i,n}$, $i=1,\ldots,K$, $n=0,\ldots,N-1$. {\ed The power of the additive noise is set according to the receiver noise figure $N_0B$, i.e., $N_0B = k_BT_0B$, $k_B$ being the Boltzmann constant and $T_0$ the standard thermal noise temperature.}

\subsubsection{System parameters} We set the observation period  $T_{obs} = T_c = 325$ $\mu$s, with $T_c$ the channel coherence time obtained from the assumed Doppler spread $B_D$.\footnote{We recall that the Doppler spread $B_D$ and coherence time $T_c$ are inversely proportional to one another (see \cite{rappaport}).} As a consequence, the number of finite samples that can be collected for each observation is $N = 16$, which corresponds to a signal bandwidth  $B$ from $25$ to $50$ kHz according to the choice of the roll-off factor $\alpha_{\text{\tiny RRCR}}$. In Appendix we report the exact computation of such values, which shows the existence of system parameter settings such that all the assumptions given before \eqref{eq::modelflatslow} (\emph{flat} and \emph{slow} fading) are jointly satisfied.

\subsubsection{Competitors} 
As concerns the competitors, we consider the {\ed SP and Max-power DPE algorithms --- which we recall can be considered as extensions to the mobile case of \cite{weiss2,weiss1} and \cite{weiss4,weiss5}, respectively --- } and a modified version of the WLS proposed in \cite{CCFR}, which is an IPE. More precisely, we replaced the MUSIC algorithm (inapplicable here) with the smooth-MUSIC, followed by the application of the beamforming to estimate the LOS direction as the angle associated with the strongest output power.

\subsection{Localization based on single Base Station}

\hspace{-0.5cm}\begin{figure}
\begin{center}
    \begin{subfigure}[b]{0.52\textwidth}
        \includegraphics[width=\textwidth]{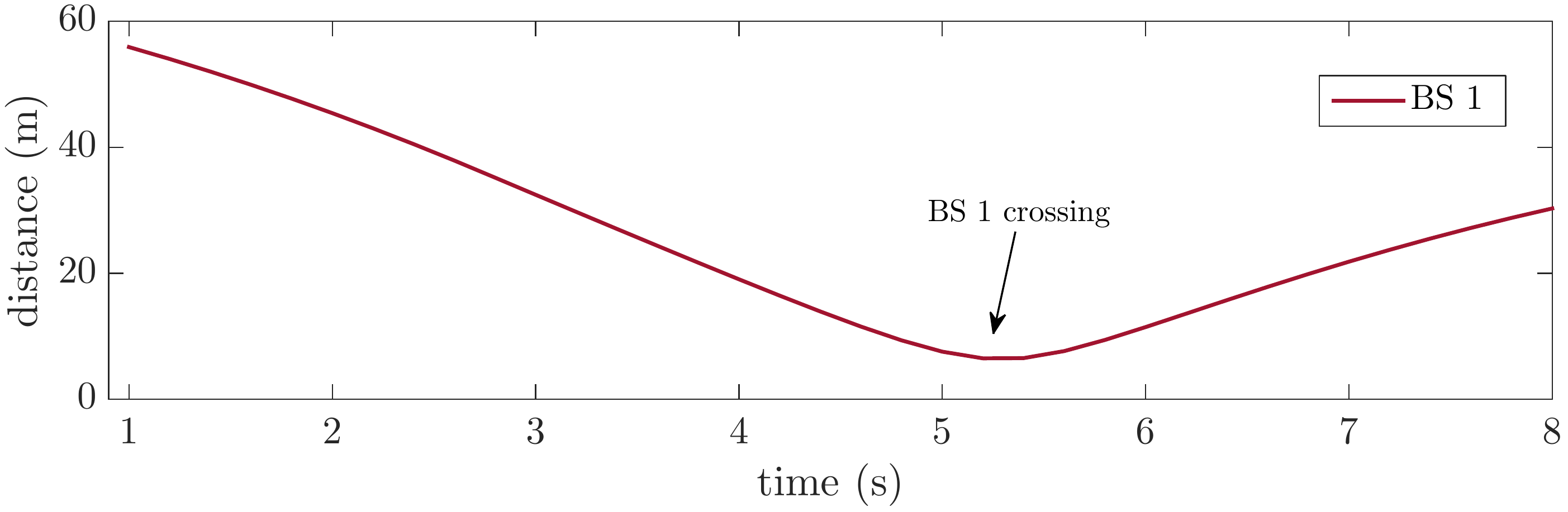}
\vspace{-0.5cm}        
        \caption{}\label{fig:dist1BS}
    \end{subfigure}
    \\
    \begin{subfigure}[b]{0.52\textwidth}
        \includegraphics[width=\textwidth]{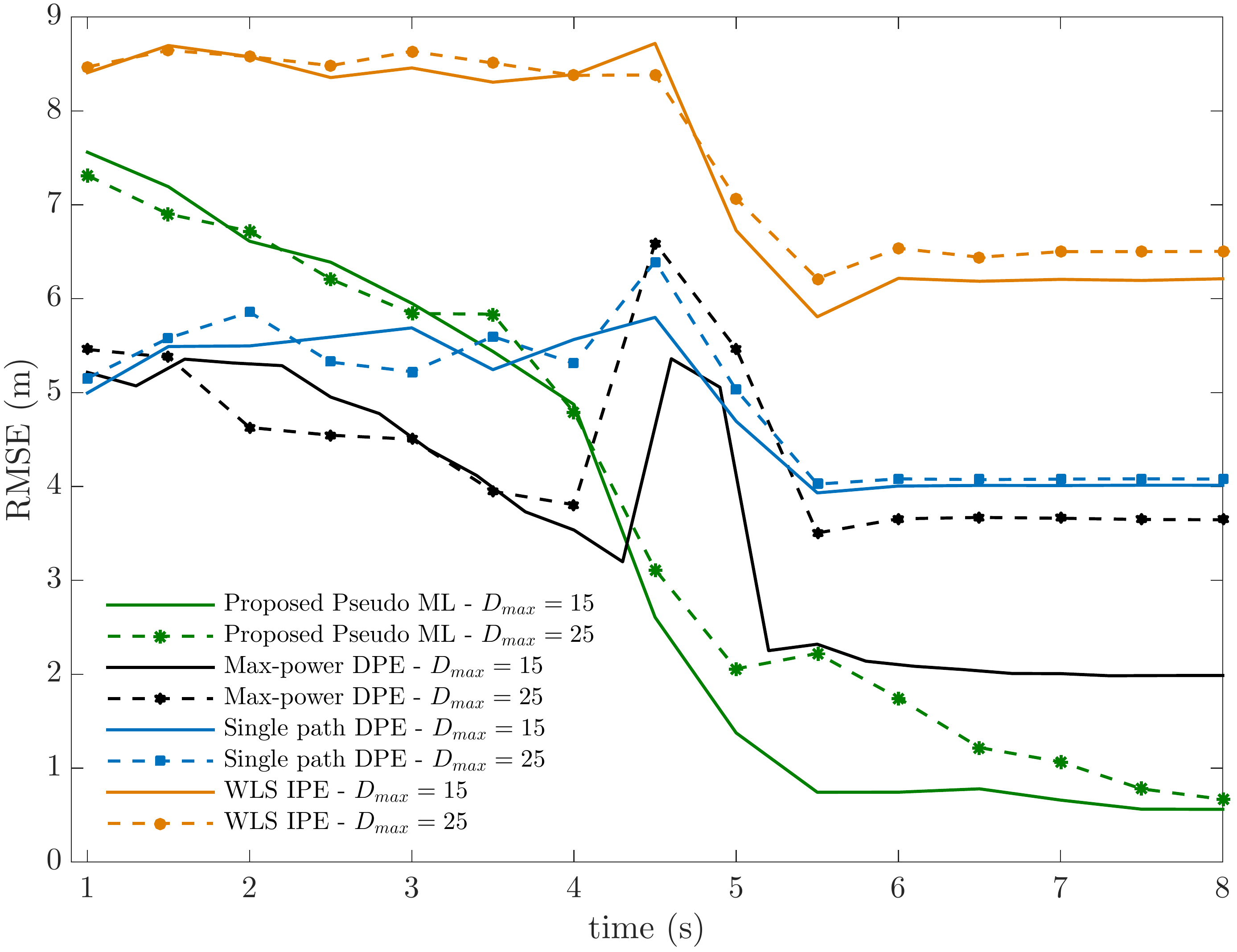}
           \caption{}\label{fig:1BS}
    \end{subfigure}
    \end{center}
    \hfill
    \vspace{-0.2cm}        
    \caption{(a) Distance between the MS and a single BS as function of the time. (b) RMSEs for the case of  $D_{\text{max}}= 15$ in comparison with the performance obtained for $D_{\text{max}} = 25$.}\label{fig:singleBS}
\end{figure}
The first analyzed scenario is a minimal situation involving only the MS and a single BS initially distant 60 m. {\ed The MS moves towards the BS and crosses it after about 5.35 s, as shown in Fig.~\ref{fig:dist1BS}.
The RMSEs of the MS position estimates for the different algorithms are reported in Fig.~\ref{fig:1BS}.
We consider two different levels of multipath: in cases of moderate multipath conditions, we set $D_{\text{max}} = 15$ while we generate the actual $D_i \sim \mathcal{U}(10,D_{\text{max}})$; in cases of more severe multipath, $D_{\text{max}} = 25$ while the actual $D_i \sim \mathcal{U}(15,D_{\text{max}})$. }
As it can be seen, the {\ed pseudo ML algorithm} exhibits an error that immediately starts to decrease as more measurements are available. Interestingly, the RMSE abruptly drops between 4 and 5 s, exhibiting values below one meter and thus outperforming all the other methods. In fact, the proposed algorithm is able to consistently exploit the additional information progressively available to correctly identify the most probable initial position $\tilde{\bm{p}}_0$, thanks to a more and more accurate reconstruction of the unknown optimal projector $\bm{P}_{\bm{x}_i}$. {\ed Notice that the performance are still remarkable also for the  challenging case of higher multipath ($D_{\text{max}}=25$, dashed lines with markers), with only a slightly longer settling time.}

The localization capability of the simpler Max-power DPE algorithm is also interesting, at least for non-severe multipath conditions. {\ed More precisely, we can observe that the RMSE tends to decrease as the MS approaches the BS (except for the short period corresponding to AOAs close to 90 degrees), stabilizing around values of error close to 2 or 4 m, depending on the multipath level. From this behavior it can be deduced that, as long as the multipath environment is not severe, the Max-power DPE can be a simpler alternative to the pseudo ML if the provided (inferior) accuracy is sufficient for the application at hand}. Conversely, the SP algorithm exhibits unsatisfactory  performance, meaning that the effects of multipath cannot be neglected in the considered scenario.

As regards the WLS IPE, the results clearly show that its performance is totally unacceptable even under milder multipath conditions, with a gap of more than 600\% compared to the proposed {\ed pseudo ML.} This result is in agreement with the generally worse performance of  IPE approaches compared to  DPE ones already observed in the literature (ref. Sec. \ref{sec:relatedwork}).

{\ed To conclude the analysis, we investigate the behavior of the proposed algorithms when the actual number of multipath components $D_i$ can exceed the assumed $D_{\text{max}}$. In particular, we consider the more challenging case of higher multipath where  $D_{\text{max}} = 25$ is assumed while the actual number $D_i$ is randomly chosen between 20 and 35. The obtained results are depicted in Fig.~\ref{fig:singleBS_exceedreplic}. 
\begin{figure}
\centering
	\includegraphics[width=0.50\textwidth]{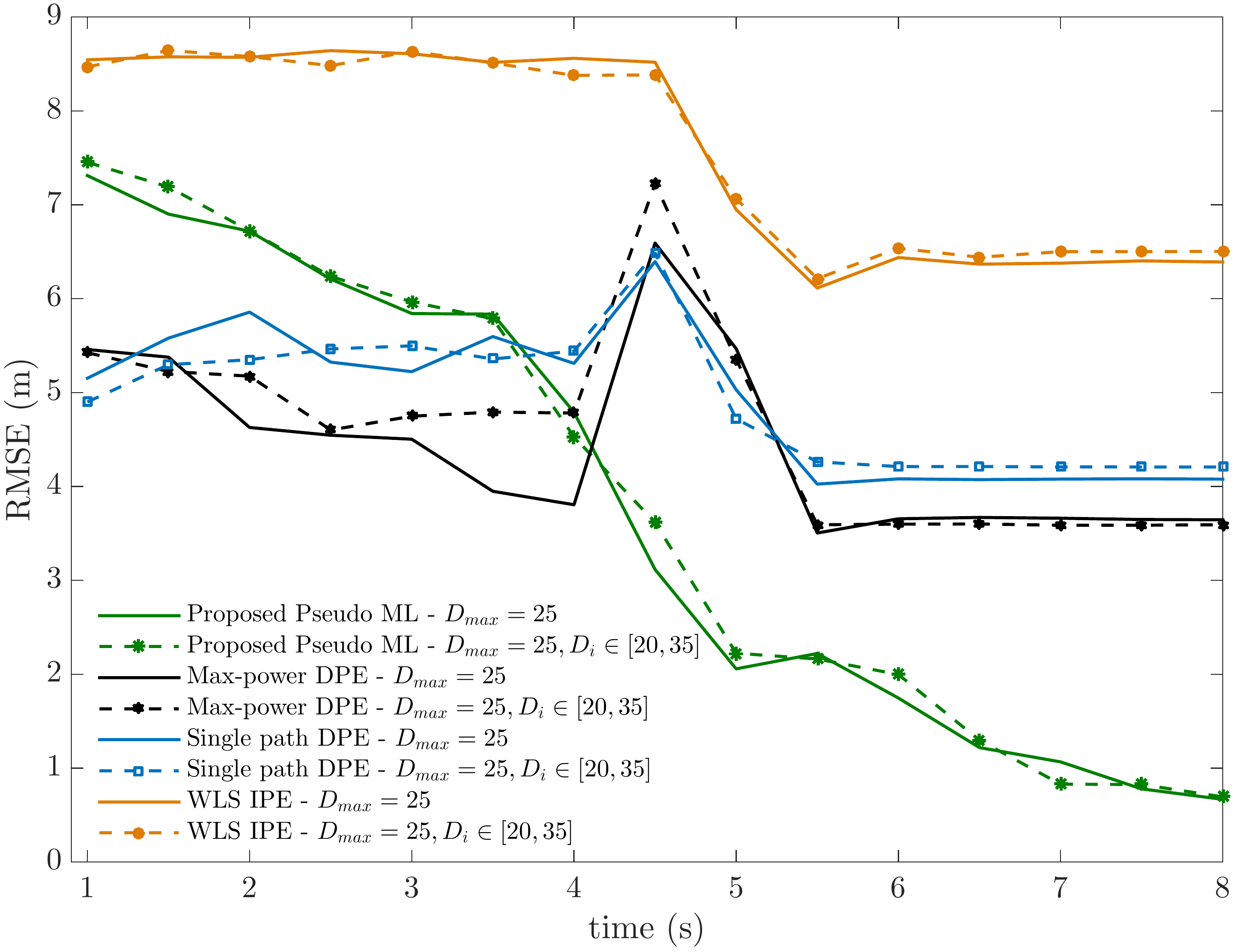}
	\caption{RMSEs for the case of $D_{\text{max}} = 25$ and  $D_i \in [20,35]$.}
	\label{fig:singleBS_exceedreplic}
\end{figure}
Remarkably, both the pseudo ML and Max-power algorithms are very robust against a non-perfect knowledge of the operating environment.}

\subsection{Localization based on multiple Base Stations}

\begin{figure}
\begin{center}
    \begin{subfigure}[b]{0.5\textwidth}
        \includegraphics[width=\textwidth]{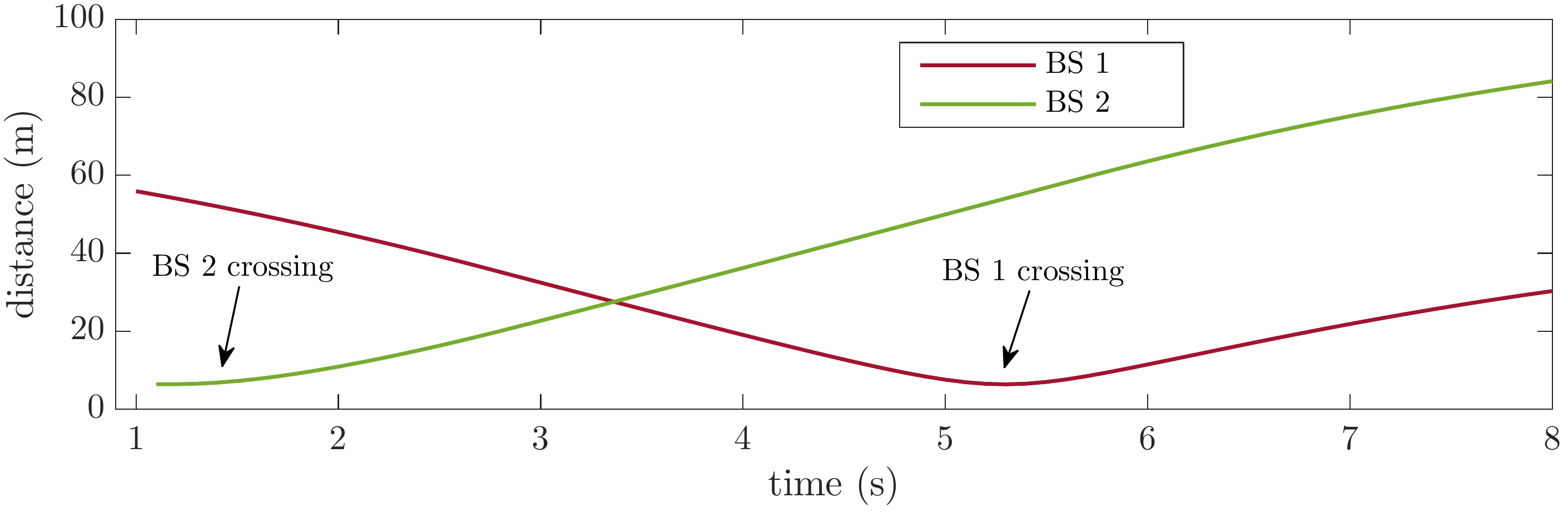}
\vspace{-0.5cm}        
        \caption{}\label{fig:dist2BS}
    \end{subfigure}
    \\
    \begin{subfigure}[b]{0.5\textwidth}
        \includegraphics[width=\textwidth]{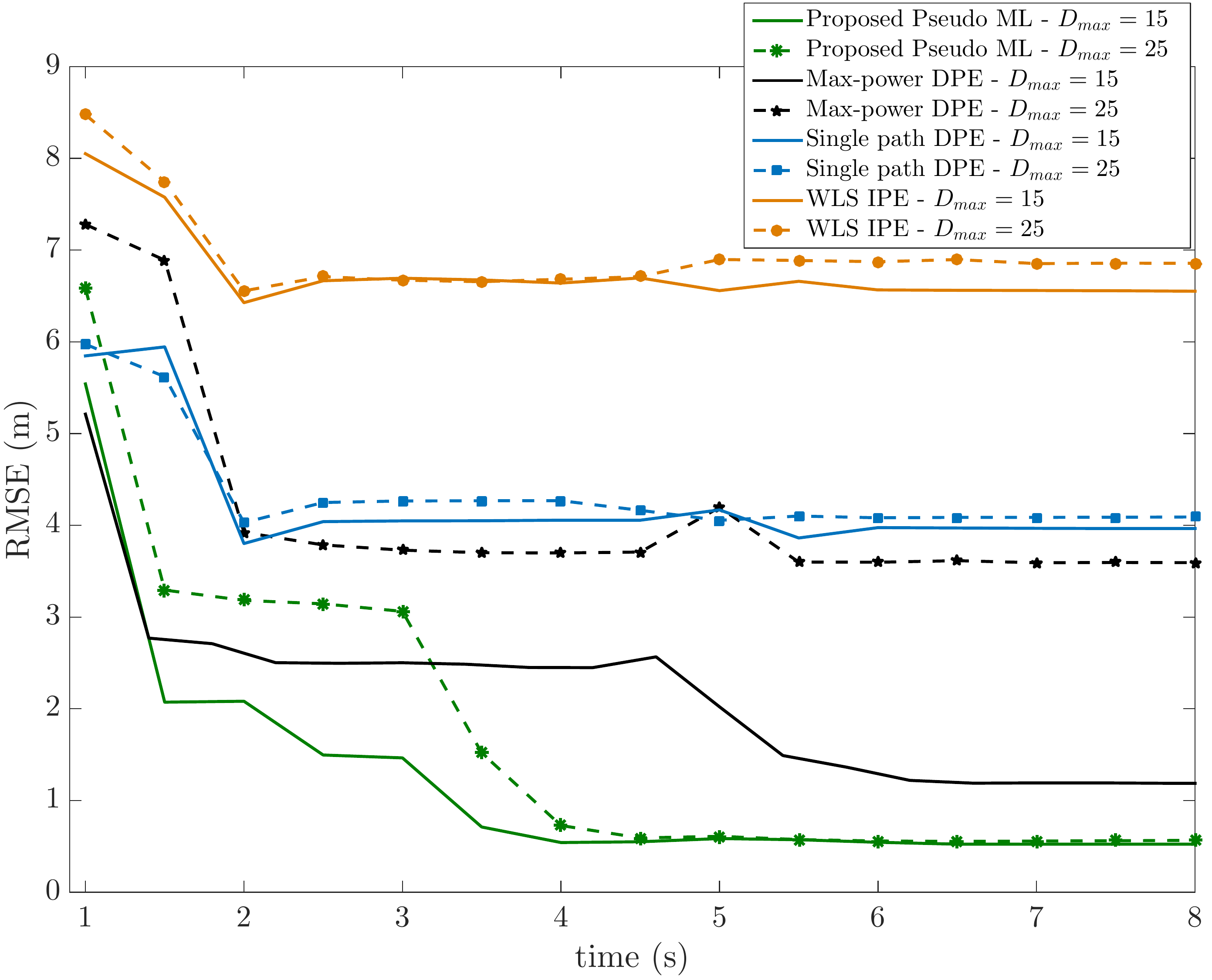}
     \caption{}\label{fig:2BS}
    \end{subfigure}
    \end{center}
    \hfill
    \vspace{-0.2cm}        
    \caption{(a) Distance between the MS and two BSs as function of the time. (b) RMSEs for the case of  $D_{\text{max}} = 15$ in comparison with the performance obtained for $D_{\text{max}} = 25$.}
\end{figure}

When more BSs are available in range, the performance generally improve but, remarkably, the gain for the proposed {\ed pseudo ML} is dramatic. Indeed, by considering just two BSs instead of one (in particular, one BS is still at 60 m while a second one is at 20 m) the RMSE immediately drops to sub-meter accuracy, also for severe multipath conditions {\ed (generated as for the case of Fig. \ref{fig:singleBS}),} as shown in Fig. \ref{fig:2BS}. {\ed Conversely, for $D_{\text{max}} = 25$ the single-path  and max-power algorithms show almost flat performance over time, meaning that, due to the severe multipath, they are not able to take advantage of the additional information collected during the motion. It is though worth noticing that for  reduced multipath the Max-power algorithm has  good performance, although it requires a  large number of measurements to attain about 1-meter accuracy.}

\begin{figure}
    \begin{subfigure}[b]{0.5\textwidth}
        \centering \includegraphics[width=0.85\textwidth]{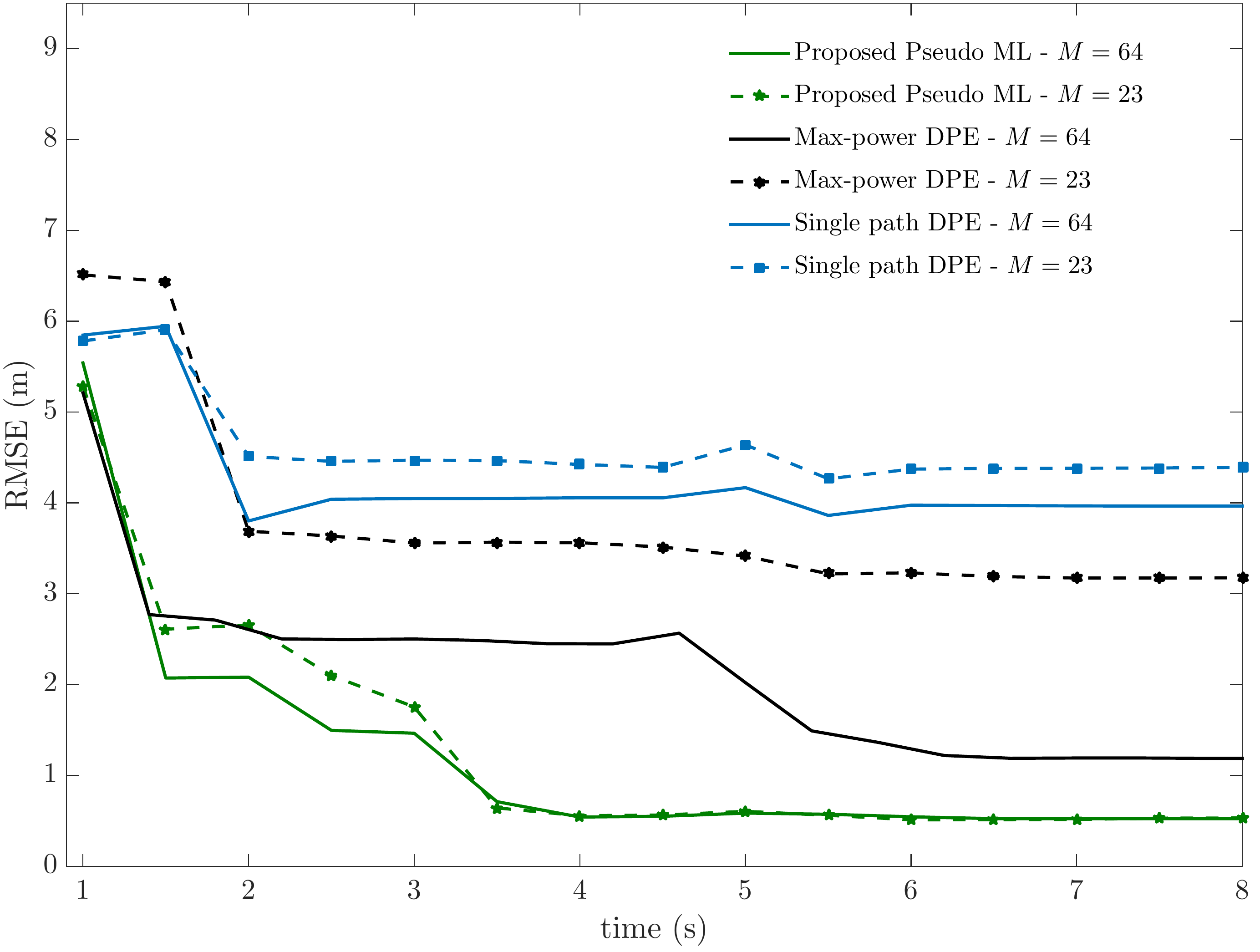}       
        \caption{}\label{fig:dist2BS_minant_10}
    \end{subfigure}
    \begin{subfigure}[b]{0.5\textwidth}
       \centering \includegraphics[width=0.85\textwidth]{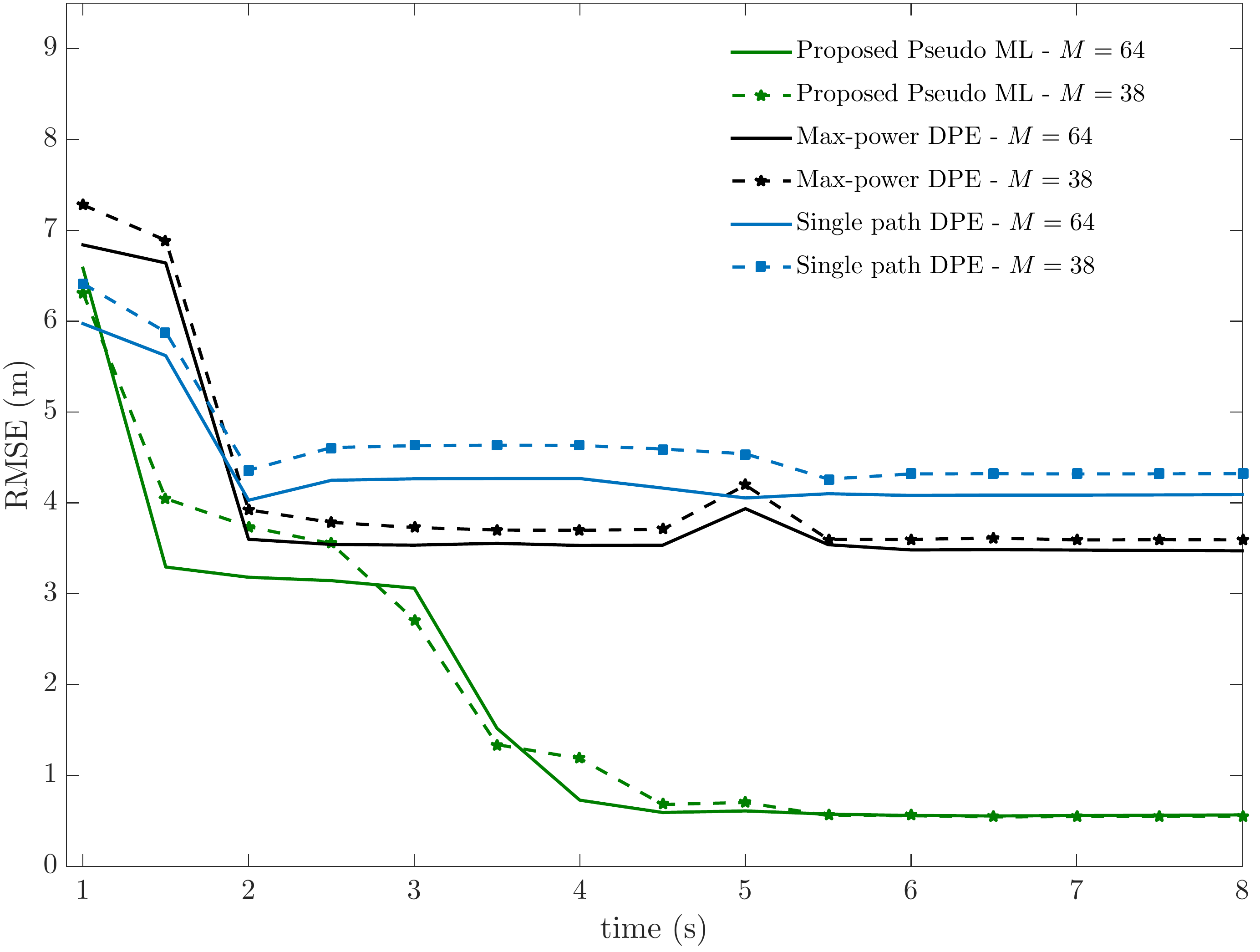}
           \caption{}\label{fig:2BS_minant_25}
    \end{subfigure}
    \hfill
    \vspace{-0.2cm}        
    \caption{(a) RMSEs for the case $D_{\text{max}} = 15$ and for both $M = 64$ and $M = 23$. (b) RMSEs for the case of  $D_{\text{max}} = 25$ and for both $M = 64$ and $M = 38$.}\label{fig:2BS_minant}
\end{figure}

In the second part of the analysis,  we show the performance of the algorithms when the minimal value for the number of antennas $M$ is considered. {\ed Specifically, given a value of $D_{\text{max}}$, we recall that the theoretical minimum number of antennas to make the problem well-posed is obtained considering $S = (D_{\text{max}}+1)/2$ subarrays, each of length $P = D_{\text{max}} +1$ \cite{vantrees}; this leads to $M = 23$ and $M = 38$ for $D_{\text{max}} = 15$ and $D_{\text{max}} = 25$, respectively.} The obtained results are reported in Fig.~\ref{fig:2BS_minant}. Remarkably, the proposed {\ed pseudo ML} algorithm still exhibits the best performance. Again, for more severe multipath, the RMSE of the competitors has a floor due to the fact that such algorithms are not able to get rid of the interfering NLOS paths.
{\ed It can be noticed that the RMSE curves of the pseudo ML algorithm exhibit a gap during the first part of the simulation, meaning that the reduced number of antennas $M$ (23 or 38 according to the case with small or large number of paths, ref. cases (a) and (b) in Fig. \ref{fig:2BS_minant}, respectively) is affecting the achieved localization performance. In the second part of the simulation, the performance tends to stabilize around values of localization errors that are comparable with those obtained in the case of $M = 64$ antennas. This behavior can be attributed to the beneficial effects of performing  LOS associations over the MS trajectory; by dynamically updating the number of LOS associations performed for each trial point $\tilde{\bm{p}}_0$ in the initial grid $\mathcal{P}_0$ over time, the pseudo ML algorithm is able to gain an increasingly more accurate belief that is used to identify a restricted set $\mathcal{P}_k \subset \mathcal{P}_0$ containing the most probable position estimates at each current instant $t_k$. Such a restricted set will be then used to optimize eq. \eqref{eq::loglike_4}, leading to an integration gain which is reflected in the decreasing RMSE.}

{\ed Finally, to further challenge the proposed algorithm, we simulated a more difficult operating environment characterized by the more severe multipath, namely $D_{\text{max}} = 25$, and LOS blockages that occur in the 50\% of cases. In Fig.~\ref{fig:50NLOS}, we report the algorithms performance in comparison with the results obtained for the case of NLOS probability equal to 10\%.  
It is interesting to observe that the proposed pseudo ML algorithm is able to attain 2-meters accuracy even when the LOS path is absent in half of the collected observations.
\begin{figure}
\centering
	\includegraphics[width=0.52\textwidth]{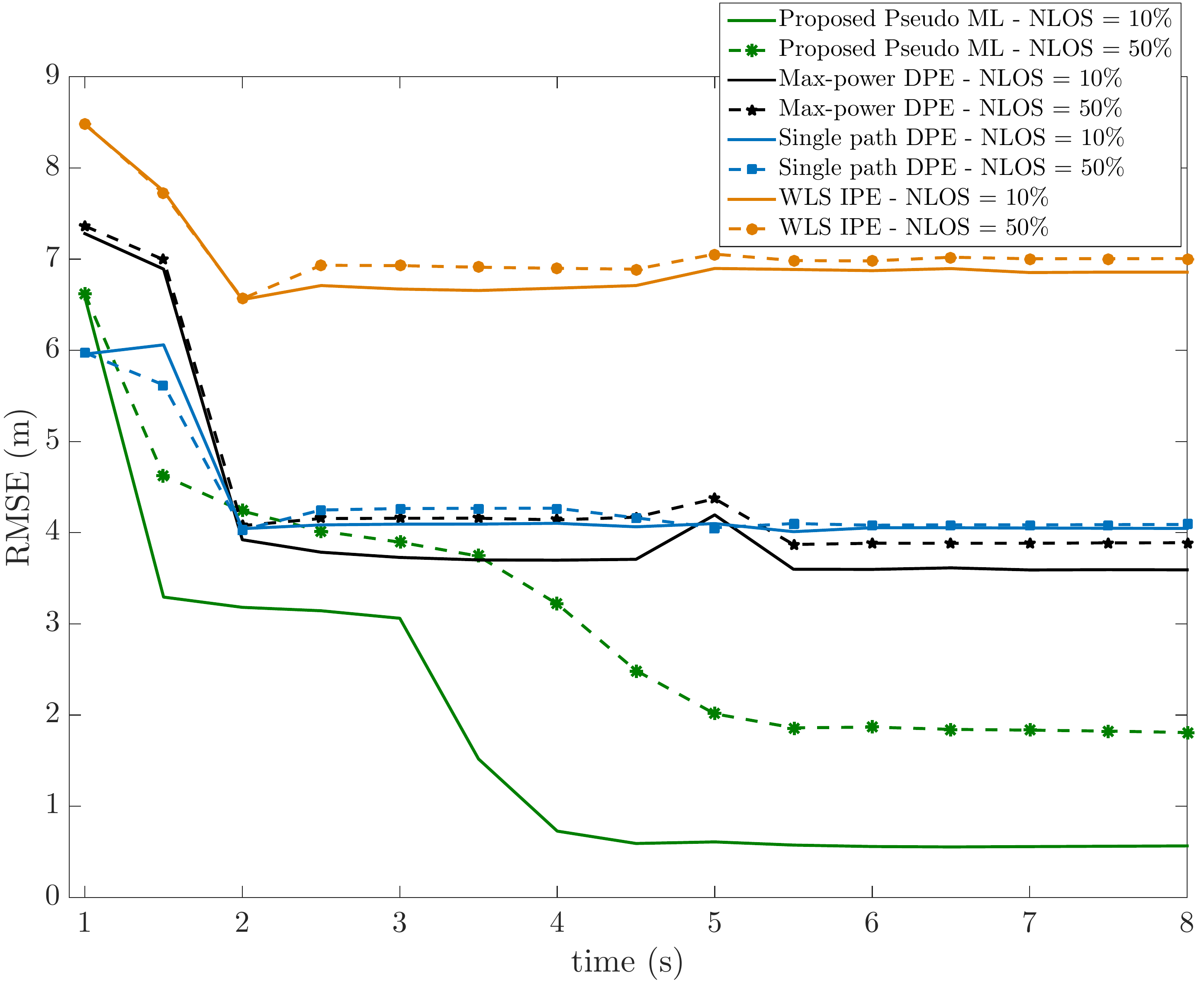}
	\caption{RMSEs  for the case of $D_{\text{max}} = 25$ and for both 10\% and 50\% of LOS blockages.}
	\label{fig:50NLOS}
\end{figure}
}
\section{Conclusion}\label{sec:conclusions}

In this work, we have designed a low complexity and fully adaptive  algorithm for localizing a mobile node in presence of  multipath. Our solution is based on the {\ed pseudo ML} approach and exploits only downlink radio signals. The algorithm employs an adaptive beamforming technique to reconstruct an estimate of the optimal projection matrices, which are then used to project the received signal {\ed onto the suitable directions that exploit the multipath structure}. Furthermore, it takes advantage of a simple and effective LOS association method to identify the most probable MS initial position, hence achieving an integration gain over time. The performance assessment has been conducted by means of simulations considering realistic values of kinematic, communication, and environmental parameters. The  results revealed that the proposed algorithm is very effective even in presence of few {\ed (or even one) BSs and severe multipath,  outperforming state-of-the-art competitors.}

\appendix

In this section, we conduct an analysis aimed at determining a possible choice of the system parameters $T_{obs}$ and $B$ such that all the assumptions before \eqref{eq::modelflatslow} are jointly satisfied.


We first  analyze the MS mobility to identify a  time interval, referred to as $T_{sta}$,  in which it is possible to  assume that the multipath environment remains practically unchanged. By constraining the variation of the steering vector   to be lower than a  threshold $\kappa$ over a finite interval $[t_i,t_i + \Delta T]$, we obtain
\begin{align}\label{eq:vincoloAOA}
T_{sta} = &\max \Delta T \nonumber \\
&s.t. \;\|\bm{a}(\theta^{\text{\tiny LOS}}_i) - \bm{a}(\theta^{\text{\tiny LOS}}_{i+\Delta T})\| \leq \kappa
\end{align} 
where $\theta^{\text{\tiny LOS}}_{i+\Delta T}$ is readily determined from the computation of $\bm{p}(t_{i+\Delta T}$) through \eqref{eq::kinmodel}, followed by the application of the geometric model in \eqref{eq::aoa}. We consider a linear trajectory at a constant velocity of 50 km/h, which is the maximum speed considered in the simulations hence represents a conservative choice. Solving the constrained problem in \eqref{eq:vincoloAOA} for $\kappa = 0.01$ allows us to determine the value of $T_{sta}$ such that the maximum variation of $\bm{a}(\theta^{\text{\tiny LOS}}_i)$ due to MS mobility does not exceed $1\%$, i.e., it is practically negligible. Notice that the entity of the variation strictly depends on the nonlinear relation between $\bm{p}(t_i)$ and $\theta^{\text{\tiny LOS}}_i$, as given in \eqref{eq::aoa}.  Therefore, two different cases have been analyzed: i) the MS is 100 m far from the BS; ii) the MS is 20 m far from the BS. The resulting value of $T_{sta}$ is 122 ms in the first case and 6 ms in the second case, respectively. As it can be noticed, the variation is much higher for closer distances, as direct consequence of the nonlinear increase of $\theta^{\text{\tiny LOS}}_i$ as the MS approaches the BS. However, the values of $T_{sta}$ are significantly greater than the channel coherence time $T_c$, meaning that this latter represents the most stringent constraint.

By jointly considering all the assumptions discussed in Sec. \ref{sec:sysmodel}, we obtain the  constraints
$ 
\left\{
\begin{array}{ll}
 &\!\!\!\!\!\!\! T_{obs} \leq T_c \quad \\ 
 & \!\!\!\!\!\!\! B \gg B_D \\
 & \!\!\!\!\!\!\! B \ll B_c(1+\alpha_{\text{\tiny RRCR}})
\end{array}
\right. .
$
\begin{figure}
\centering
	\hspace{-0.4cm}\includegraphics[width=0.5\textwidth]{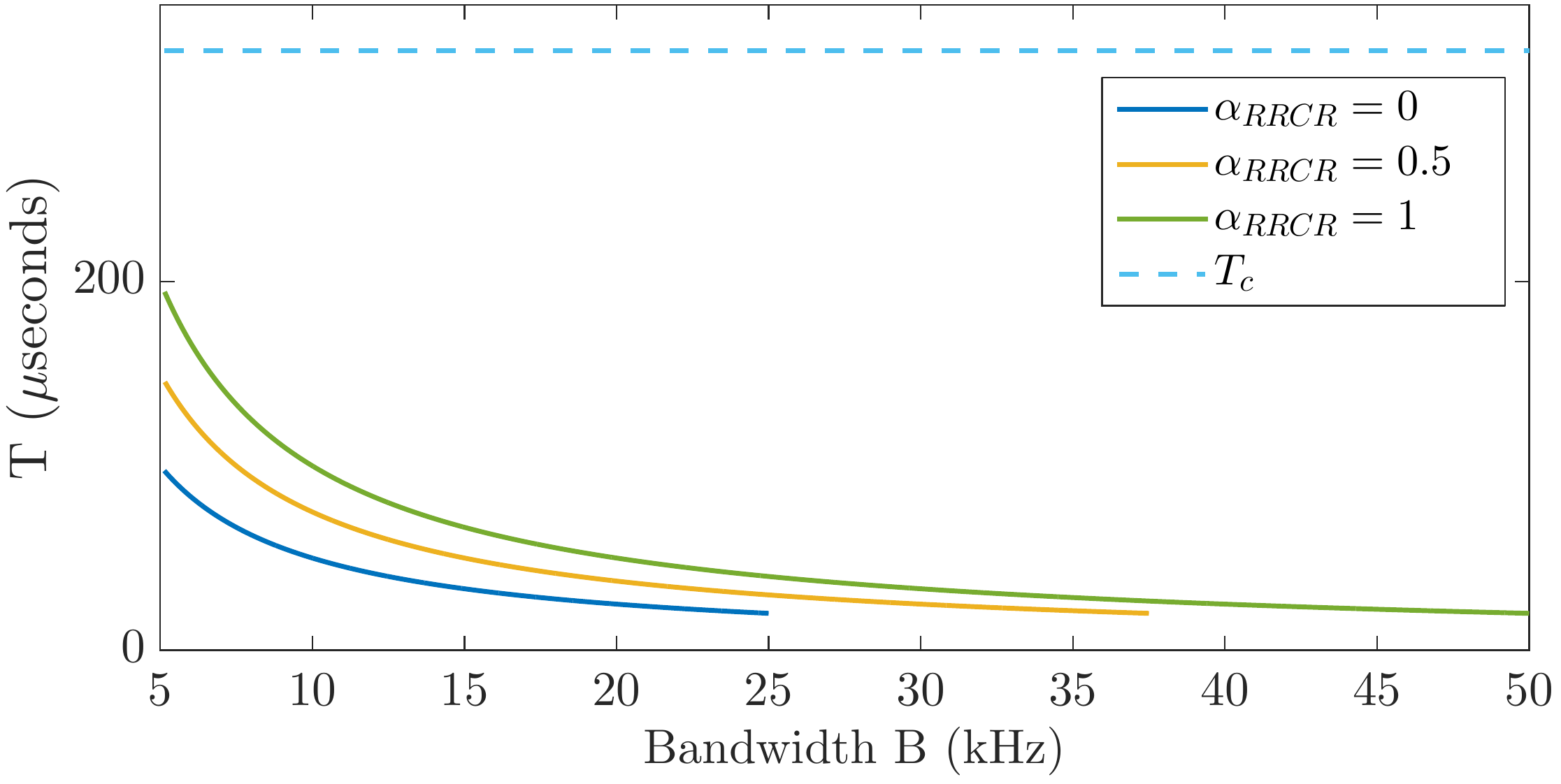}
	\caption{Sampling interval $T$ as function of the bandwidth $B$ for three different values of the roll-off factor $\alpha_{\text{\tiny RRCR}}$.}
	\label{fig:maxsnapflatslow}
\end{figure}
In Fig.~\ref{fig:maxsnapflatslow} we draw the values of the sampling interval  $T$ as function of the bandwidth $B$ satisfying all the conditions above. 
As it can be observed, $T$ is much lower than the channel coherence time $T_c$ for all the possible values of roll-off $\alpha_{\text{\tiny RRCR}}$. Moreover, it should be noticed that different values of $\alpha_{\text{\tiny RRCR}}$ give rise to different ranges of allowed $B$, as shown by the three solid curves. Interestingly, all the three hyperbolas attain the same minimum value of $T$, leading to the same number of collected samples $N = \left\lfloor\frac{T_{obs}}{T} \right\rfloor = 16$, but for different bandwidths $B$. Thus, standing the same value of $N$, one can consider the most convenient choice of $\alpha_{\text{\tiny RRCR}}$ based on practical considerations regarding bandwidth consumption and ease of electronic implementation of the pulse shaper. 


\end{document}